\documentclass[twocolumn,showpacs,preprintnumbers,amsmath,amssymb,footinbib]{revtex4}

\usepackage{graphicx}
\usepackage{dcolumn}
\usepackage{bm}
\usepackage{epsfig}
\usepackage{epstopdf}
\usepackage{enumerate}
\usepackage{mathrsfs}

\begin{document}

\preprint{APS/123-QED}

\title{Third-order spontaneous parametric downconversion in thin optical fibers as a photon-triplet source}

\author{Mar\'ia Corona,$^{1,2,*}$ Karina Garay-Palmett,$^1$ and Alfred B. U'Ren$^{1}$}
\affiliation{$^1$Instituto de Ciencias Nucleares, Universidad Nacional Aut\'onoma de M\'exico, apdo. postal 70-543, M\'exico 04510 DF
\\
$^2$Departamento de \'Optica, Centro de Investigaci\'on Cient\'{\i}fica y
de Educaci\'on Superior de Ensenada, Apartado Postal 2732, Ensenada,
BC 22860, M\'exico \\
$^*$Corresponding author: maria.corona@nucleares.unam.mx
}

\date{\today}

\newcommand{\epsfg}[2]{\centerline{\scalebox{#2}{\epsfbox{#1}}}}

\begin{abstract}
We study the third-order spontaneous parametric downconversion
(TOSDPC) process, as a means to generate entangled photon triplets.
Specifically, we consider thin optical fibers as the nonlinear
medium to be used as the basis for TOSPDC, in configurations where
phasematching is attained through the use of different fiber
transverse modes. Our analysis in this paper, which follows from our
earlier paper Opt. Lett.  \textbf{36}, 190--192 (2011), aims to
supply experimentalists with the details required in order to design
a TOSPDC photon-triplet source.  Specifically, our analysis focuses
on the photon triplet state, on the rate of emission, and on the
TOSPDC phasematching characteristics, for the cases of
frequency-degenerate and frequency non-degenerate TOSPDC.
\end{abstract}

\maketitle


\section{Introduction}

The generation of entangled photon multiplets represents an
important goal in quantum optics, as a resource for fundamental
tests of quantum mechanics as well as for the implementation of
quantum-enhanced technologies. A large number of experiments from
the last few decades have exploited entangled photon \emph{pairs}
generated by the process of spontaneous parametric downconversion
(SPDC) in second-order non-linear crystals~\cite{burnham70}.
Recently, the process of spontaneous four wave mixing (SFWM) based
on the third-order nonlinearity of optical fibers has emerged as a
viable alternative to SPDC for the generation of photon
pairs~\cite{fiorentino02}.   However, the generation of entangled
photon \emph{triplets}, and of higher-order entangled photon
multiplets, faces acute technological challenges.

The motivation which served as starting point for the present work
is that in principle the same third order non-linearity in fused
silica optical fibers which is responsible for the SFWM process also
permits a different process: third-order spontaneous parametric
downconversion (TOSPDC)~\cite{chekhova05,hnilo05,felbinger98,
bencheikh07,banaszek97,douady04}.   While in the SFWM process, two
pump photons are jointly annihilated in order to generate a photon
\emph{pair}, in the TOSPDC process a single pump photon is
annihilated in order to generate a photon \emph{triplet}. TOSPDC may
be differentiated from other approaches based on nonlinear optics
for the generation of photon triplets, by the fact that the three
photons in a given triplet are derived from a single
quantum-mechanical event.  The prospect of efficient generation of
photon triplets is exciting on a number of fronts.  On the one hand,
it naturally leads to the possibility of heralded emission of photon
pairs~\cite{Sliwa03,Wagenknecht10,Barz10}.  On the other hand, it
leads to the possibility of direct generation of
Greenberger-Horne-Zeilinger (GHZ) polarization-entangled
states\cite{greenberger90,bouwmeester03}, without resorting to
postselection. In addition, if the photon triplets are emitted in a
single transverse-mode environment they exhibit factorability in
transverse momentum, but can exhibit spectral entanglement. Such
three-partite entanglement in a continuous degree of freedom is a
potentially important, yet largely unexplored topic.

A number of approaches for the generation of photon triplets have
been proposed including:  i) triexcitonic
decay in quantum dots~\cite{persson04}, ii) combined, or cascaded,
second-order nonlinear
processes~\cite{keller98,hubel10,antonosyan11}, and iii)
approximate photon triplets formed by SPDC photon pairs together
with an attenuated coherent state~\cite{rarity99}.  Of these approaches,
those that have been experimentally demonstrated lead to very low photon-triplet detection rates.
Recently, we have proposed a specific technique for the generation
of photon triplets based on the TOSDPC process in thin optical
fibers and relying on multiple transverse fiber
modes~\cite{corona11}. As will be discussed below, the emission
rates predicted for a source based on our proposal are likewise low.
However, future advances in optical fiber technology, specifically
in the form of highly-nonlinear fibers, photonic crystal fibers, and
tapered fibers may significantly increase the emitted
flux attainable through our proposal.

The purpose of this paper is to explore the theory behind  our
proposal for TOSPDC photon-triplet sources.  In particular, we focus
on the photon-triplet state,  on the rate of emission, and on the
TOSPDC phasematching characteristics of thin optical fibers.   In
order to make our analysis as general as possible, we include both
frequency-degenerate and frequency non-degenerate TOSPDC, as well as
both the monochromatic- and pulsed-pumped regimes.

\section{Derivation of the photon-triplet quantum state}

In this paper we study the process of third-order spontaneous
parametric downconversion (TOSPDC) in optical fibers, in which
nonlinear phenomena originate from the third-order electrical
susceptibility $\chi^{(3)}$. In this process, individual photons
from the pump mode (p), may be annihilated giving rise to the
emission of a photon triplet. Borrowing from second-order
spontaneous parametric downconversion terminology, we refer to the
three emission modes as signal-$1$ (r), signal-$2$ (s), and idler
(i). We restrict our analysis to configurations for which the three
TOSPDC photons are generated in the same transverse fiber mode, and
where all four fields are co-polarized (with linear polarization
along the $x$-axis) propagating in the same direction along the
fiber (which defines the $z$-axis).

It can be shown that the TOSPDC process is governed by the following
Hamiltonian

\begin{eqnarray}\label{Eq:hamiltonian}
    \hat{H}(t)&=&\frac{3}{4}\epsilon_0 \chi^{(3)}\nonumber\\
    &\times&\int dV \hat{E}_p^{(+)}(\bold{r},t)\hat{E}_{r}^{(-)}(\bold{r},t)\hat{E}_{s}^{(-)}(\bold{r},t)\hat{E}_i^{(-)}(\bold{r},t),\nonumber\\
\end{eqnarray}

\noindent in terms of the positive-frequency and negative-frequency parts
of the electric field operator (denoted by (+)/(-) superscripts) for
each of the modes, labeled as $\mu=p,r,s,i$.  In
Eq.~(\ref{Eq:hamiltonian}), $\epsilon_0$ represents the vacuum
electric susceptibility, and the integral is evaluated over the
nonlinear medium volume illuminated by the pump field.
$E_{\mu}^{(+)}(\bold{r},t)$ (with $\mu=r,s,i$) may be written as

\begin{equation}\label{Eq:photonfields}
    \hat{E}^{(+)}(\bold{r},t)=iA(x,y)\sqrt{\delta k}\sum_{k}\ell(\omega)\exp[i(k z-\omega t)]\hat{a}(k),
\end{equation}

\noindent where $\hat{a}(k)$ is the wavenumber-dependent
annihilation operator associated with the propagation mode in the
fiber, and $\delta k=2\pi/L_Q$ is the mode spacing defined in terms
of the quantization length $L_Q$.  $A(x,y)$ represents the
transverse spatial distribution of the field, which is approximated
to be frequency-independent within the bandwidth of the generated
wave-packets, and is normalized so that $\int\int|A(x,y)|^2dxdy=1$.
In Eq.~(\ref{Eq:photonfields}) the function $\ell[\omega(k)]$ is
given as

\begin{equation}\label{Eq:ll}
    \ell(\omega)=\sqrt{\frac{\hbar\omega}{\pi\epsilon_0 n^{2}(\omega)}},
\end{equation}

\noindent where $n(\omega)$ is the refractive index of the medium
and $\hbar$ is Planck's constant.

For the analysis presented here, we describe the pump mode as a
classical field, expressed in terms of its Fourier components as

\begin{equation}\label{Eq:Ep}
    E_p^{(+)}(\bold{r},t)=A_0A_p(x,y)\int d\omega_p\alpha(\omega_p)\exp[i(k_p(\omega_p) z-\omega_p t)],
\end{equation}

\noindent in terms of the pump-mode amplitude $A_0$, and the pump
transverse distribution in the fiber $A_p(x,y)$,  normalized so that
$\int\int|A_p(x,y)|^2 dx dy=1$, and approximated to be
frequency-independent within the pump bandwidth.  In
Eq.~(\ref{Eq:Ep}), the function $\alpha(\omega_p)$ is the pump
spectral amplitude (PSA), with normalization
$\int|\alpha(\omega)|^2d\omega=1$. It can be shown that $A_0$ is
related to pump peak power $P$ through

\begin{equation}
    A_0=\sqrt{\frac{2P}{\epsilon_0 c n_p|\int d\omega_p\alpha(\omega_p)|^{2}}},
\end{equation}

\noindent where $n_p\equiv n(\omega_{p0})$; $\omega_{p0}$ is the
pump carrier frequency.

By replacing Eqs.~(\ref{Eq:photonfields}) and (\ref{Eq:Ep}) into
Eq.~(\ref{Eq:hamiltonian}), and following a standard perturbative
approach~\cite{mandel}, it can be shown that the state produced by
third-order spontaneous parametric down conversion is
$|\Psi\rangle=|0\rangle_r|0\rangle_s|0\rangle_i+\xi|\Psi_3\rangle$,
written in terms of the three-photon component of the state
$|\Psi_3\rangle$

\begin{eqnarray}\label{Eq:photonstate}
    |\Psi_3\rangle&=&\sum_{k_r}\sum_{k_s}\sum_{k_i}G_k(k_r,k_s,k_i)\nonumber\\
    &\times&\hat{a}^{\dagger}(k_r)\hat{a}^{\dagger}(k_s)\hat{a}^{\dagger}(k_i)|0\rangle_r|0\rangle_s|0\rangle_i,
\end{eqnarray}

\noindent where $\xi$, related to the conversion efficiency,
is given by

\begin{eqnarray}
    \xi &=& i\frac{3\epsilon_0\chi^{(3)}(2\pi)A_0(\delta k)^{3/2}L}{4\hbar}\nonumber\\
    &\times&\int dx\int dy A_p(x,y)A_r^{*}(x,y)A_s^{*}(x,y)A_i^{*}(x,y).
\end{eqnarray}

In Eq.~(\ref{Eq:photonstate}), $G_k(k_r,k_s,k_i)$ is the wavenumber
joint amplitude.  Writing this function in terms of frequencies
leads to
$G(\omega_r,\omega_s,\omega_i)=\ell(\omega_r)\ell({\omega_s})\ell(\omega_i)F(\omega_r,\omega_s,\omega_i)$.
The function $\ell(\omega)$ has a slow dependence on frequency [see
Eq.~(\ref{Eq:ll})] over the spectral range of interest.  If this
dependence is neglected, the photon-triplet spectral properties are
fully determined by the function $F(\omega_r,\omega_s,\omega_i)$,
which from this point onwards we refer to as the joint spectral
amplitude.  It can be shown that this function can be written in
terms of the pump spectral amplitude (PSA)  $\alpha(\omega)$, and
the phasematching function (PM)  $\phi(\omega_r,\omega_s,\omega_i)$
as

\begin{equation}\label{Eq:FF}
F(\omega_r,\omega_s,\omega_i)=\alpha(\omega_r+\omega_s+\omega_i)\phi(\omega_r,\omega_s,\omega_i),
\end{equation}

\noindent with

\begin{eqnarray} \label{fphmat}
\phi(\omega_r,\omega_s,\omega_i)&=&\mbox{sinc}\left[L\Delta
k(\omega_r,\omega_s,\omega_i)/2\right]\nonumber\\
&\times&\exp[iL\Delta k(\omega_r,\omega_s,\omega_i)/2],
\end{eqnarray}

\noindent written in turn in terms of the fiber length $L$ and the
phasemismatch $\Delta k(\omega_r,\omega_s,\omega_i)$

\begin{eqnarray}\label{Eq:PM}
    \Delta k(\omega_r,\omega_s,\omega_i)&=&
    k_p(\omega_r+\omega_s+\omega_i)-
    k_r(\omega_r)-k_s(\omega_s)\nonumber\\&-&k_i(\omega_i)+\Phi_{NL}.
\end{eqnarray}

In Eq.~(\ref{Eq:PM}), the last term is a non-linear contribution
written as
$\Phi_{NL}=[\gamma_p-2(\gamma_{pr}+\gamma_{ps}+\gamma_{pi})]P$,
where $\gamma_p$ and $\gamma_{p\mu}$ are the nonlinear coefficients
derived from self-phase and cross-phase modulation, respectively
\cite{agrawal07}. These coefficients may be written as

\begin{equation}  \label{Eq:gam1}
    \gamma_p=\frac{3\chi^{(3)}\omega_{p0}}{4\epsilon_0 c^{2}n_{p}^{2}A_{eff}^{(p)}},
    \end{equation}

and

\begin{equation} \label{Eq:gam2}
    \gamma_{p\mu}=\frac{3\chi^{(3)}\omega_{\mu 0}}{4\epsilon_0 c^{2}n_{p}n_{\mu
    0}A_{eff}^{(p\mu)}},
\end{equation}

\noindent in terms of the definition $n_{\mu 0} \equiv
n_{\mu}(\omega_{\mu 0})$, where $\omega_{\mu 0}$ is the central
frequency of the generated wave-packet ($\mu=r,s,i$).
$A_{eff}^{(p)}$ and $A_{eff}^{(p\mu)}$ represent the effective
interaction areas,  given by $A_{eff}^{(p)}=\left(\int\int dx dy
|A_p(x,y)|^{4}\right)^{-1}$ and $A_{eff}^{(p\mu)}=\left(\int\int dx
dy |A_p(x,y)|^{2}|A_{\mu}(x,y)|^{2}\right)^{-1}$, respectively. Note
that these expressions for interaction areas take into account the
normalization used for the transverse spatial distributions of the
four modes.

\section{Emitted flux in the process of TOSPDC}

In what follows, we focus on calculating the emission rate of a
photon-triplet source based on the TOSPDC process. In order to
facilitate this calculation, we assume that pump photons are
suppressed through appropriate filtering at the end of the TOSPDC
fiber, so that no further photon triplets are generated beyond this
point.  For our purposes, the source brightness is defined as the
number of single photons detected in one of the three generation
modes (e.g. the signal-$1$ mode) per unit time.   For the state in
Eq.~(\ref{Eq:photonstate}), which assumes a pulsed pump, we are
specficially interested in the number of signal-$1$ single photons
emitted per pump pulse, $N_r$.  An implicit
assumption in this definition is that the photon triplets may be
split into separate spatial modes; note that this can be achieved
deterministically if the three emission modes are spectrally
non-degenerate, and can be achieved only non-deterministically if
the three modes are spectrally degenerate.  $N_r$ is given by

\begin{equation}\label{Eq:numph}
N_r=\sum_{k_r}\langle\Psi_3|\hat{a}^{\dag}(k_r)\hat{a}(k_r)|\Psi_3\rangle.
\end{equation}

Note that under ideal detection efficiency conditions, the quantity
$N_r$ also corresponds to the number of photon triplets emitted per pump pulse.
Replacing Eq.~(\ref{Eq:photonstate}) in
Eq.~(\ref{Eq:numph}) it can be shown that

\begin{equation}\label{Eq:numph1}
    N_r=\upsilon\int\!\! dk_r\!\int\!\!dk_s\!\int\!\! dk_i\,\ell^2(k_r)\ell^2(k_s)\ell^2(k_i)
     |F(k_r,k_s,k_i)|^2,
\end{equation}

\noindent where the parameter $\upsilon$ is given as
$\upsilon=(3)^{2}|\xi|^{2}/(\delta k)^{3}$. Note that because
$|\xi|^{2}$ is cubic in $\delta k$, $\upsilon$ is constant with
respect to $\delta k$, and is explicitly given by

\begin{equation} \label{Eq:cte1}
    \upsilon = \frac{2(3)^{2}(2\pi)^{2}\epsilon_0^{3}c^{3}n_p^{3}}{\hbar^{2}\omega_{p0}^{2}}\frac{\gamma^{2}L^{2}P}{|\int d\omega_p\alpha(\omega_p)|^{2}},
\end{equation}

\noindent where $\gamma$ is the nonlinear coefficient that governs
the TOSPDC process, given by

\begin{equation}     \label{Eq:gam3}
    \gamma=\frac{3\chi^{(3)}\omega_{p0}}{4\epsilon_{0}c^{2}n_p^{2}A_{eff}},
\end{equation}

\noindent where $A_{eff}$ is the effective interaction area among the
four fields, expressed as

\begin{equation}\label{E:Aeff}
    A_{eff}=\frac{1}{\int dx\int dy A_p(x,y) A_r^{*}(x,y)A_s^{*}(x,y)A_i^{*}(x,y)}.
\end{equation}

In writing Eq.~(\ref{E:Aeff}), we have taken into account the
normalization used for the transverse spatial distribution of the
four fields involved. Note that $\gamma$ is distinct from $\gamma_p$
and $\gamma_{p \mu}$ defined in Eqs.~(\ref{Eq:gam1}) and
(\ref{Eq:gam2}).

In calculating the signal-$1$-mode photon number, see
Eq.~(\ref{Eq:numph1}), $k$-vector sums have been replaced by
integrals, i.e. $\delta k\sum_k\longrightarrow\int dk$, which is
valid in the limit $L_Q\longrightarrow\infty$.

\subsection{Expressions for the emitted flux in integral form}

We begin this section with a discussion of the pulsed-pump regime.
We limit our treatment to pump fields with a Gaussian spectral envelope, which can be written in the form

\begin{equation}\label{Eq:alphaGauss}
    \alpha(\omega_p)=\frac{2^{1/4}}{\pi^{1/4}\sqrt{\sigma}}\mathsf{e}^{-\frac{(\omega_p-\omega_{p0})^{2}}{\sigma^{2}}},
\end{equation}

\noindent given in terms of the pump central frequency $\omega_{p0}$
and the pump bandwidth $\sigma$.  The number of signal-$1$-mode photons $N_r$
 resulting from an isolated pump pulse can be obtained by
replacing Eqns.~(\ref{Eq:ll}), (\ref{Eq:FF}), (\ref{Eq:cte1}) and
(\ref{Eq:alphaGauss}) into Eq.~(\ref{Eq:numph1}). We further assume that the pump mode
is in the form of a pulse train with a repetition rate $R$. Thus, the number of
signal-$1$-mode photons generated per second is given by
$N=N_rR$, from which it can be shown that

\begin{eqnarray} \label{Eq:NumFot}
N &=& \frac{2^{5/2}3^2\hbar
c^3n_p^3}{\pi^{5/2}\omega^2_{p0}}\frac{L^2\gamma^2 p}{\sigma} \!\!
\int\!\!d\omega_{r}\!\!\int\!\!d\omega_{s}
\!\!\int\!\!d\omega_i\frac{k_r'\omega_r}{n_r^2} \nonumber
\\&\times&\frac{k_s'\omega_s}{n_s^2}
\frac{k_i'\omega_i}{n_i^2} |f(\omega_{r},\omega_{s},\omega_i)|^2,
\end{eqnarray}

\noindent where $p$ is the average pump power that is related to the
peak pump power $P$ through the relation $P=p\sigma/(\sqrt{2\pi}R)$.
In the derivation of Eq.~(\ref{Eq:NumFot}), integrals over $k_r$,
$k_s$ and $k_i$ were transformed into frequency integrals through
the relationship $dk_{\mu}=k_{\mu}'d\omega_{\mu}$, where $k_{\mu}'$
represents the first frequency derivative of $k(\omega)$, evaluated
at $\omega_\mu$. The new function
$f(\omega_r,\omega_s,\omega_i)=(\pi\sigma^{2}/2)^{1/4}F(\omega_r,\omega_s,\omega_i)$,
is a version of the joint spectral amplitude
$F(\omega_r,\omega_s,\omega_i)$  [see Eq.~(\ref{Eq:FF})], which does
not contain factors in front of the exponential and sinc functions,
so that all pre-factors terms appear explicitly in
Eq.~(\ref{Eq:NumFot}).

From Eq.~(\ref{Eq:NumFot}) we can see that if the pump-power
dependence of the phasemismatch can be neglected,  the emitted flux
has a linear dependence on the pump power, which implies that the
conversion efficiency in the TOSPDC process is constant with respect
to this experimental parameter.  For sufficiently large pump powers,
there may be a deviation from this stated behavior, due to the
pump-power dependence of the phasemismatch.  The linear dependence
of the emitted flux vs pump power can be directly contrasted with
the corresponding behavior observed for the SFWM process, for which
the emitted flux is proportional to the square of the pump power
\cite{garay10}.  Because of this important difference,
photon-triplet sources based on TOSPDC are, for sufficiently high
pump powers, significantly less bright than comparable SFWM sources.
On the other hand, as in the case of SFWM, $N$ varies quadratically
with the nonlinear coefficient $\gamma$, which implies that the
emitted flux has an inverse fourth power dependence on the
transverse mode radius.  The dependence of the emitted flux on other
experimental parameters will be discussed in Sec.~\ref{flux_design}.

In order to proceed with our analysis, we define the conversion efficiency as
$\eta \equiv N/N_p$, where $N_p$ is the number of pump photons per second. For a sufficiently
narrow pump bandwidth,
$N_p$ is given by $N_p=U_pR/(\hbar\omega_{p0})$, with $U_p$ the
pulse energy. For a pump pulse with a spectral envelope given by
Eq.~(\ref{Eq:alphaGauss}) we obtain that

\begin{equation}
\label{Npump} N_p=\frac{p}{\hbar\omega_{p0}}.
\end{equation}

The triplet-photon conversion efficiency can then be written as

\begin{eqnarray}\label{Eq:conversioneff1}
    \eta &=& \frac{2^{5/2}3^{2}c^{3}\hbar^{2}n_p^{3}}{(\pi)^{5/2}\omega_{p0}}\frac{L^{2}\gamma^{2}}{\sigma}\int d\omega_r\int d\omega_s\int d\omega_i \frac{k_r'\omega_r}{n_r^2}\nonumber\\
    &\times&\frac{k_s'\omega_s}{n_s^2}
    \frac{k_i'\omega_i}{n_i^2}|f(\omega_r,\omega_s,\omega_i)|^{2}.
\end{eqnarray}

Let us now turn our attention to the monochromatic-pump limit of the
TOSPDC conversion efficiency.  It can be shown that by taking the $\sigma\rightarrow0$ limit  of Eq.~(\ref{Eq:NumFot}), the number
of photon triplets emitted per second becomes

\begin{eqnarray}\label{Ncw}
    N_{cw} &=&\frac{2^{2}3^{2}\hbar c^{3}n_p^{3}\gamma^{2}L^{2}p}{\pi^2\omega_p^{2}}\nonumber\\
    &\times&\int\!\! d\omega_r\!\! \int\!\! d\omega_s\, h(\omega_r,\omega_s,\omega_{p}-\omega_r-\omega_s)\nonumber\\
    &\times&\mbox{sinc}^{2}\left[\frac{L}{2}\Delta k_{cw}(\omega_r,\omega_s)\right],
\end{eqnarray}

\noindent while by taking the $\sigma\rightarrow0$ limit of Eq.~(\ref{Eq:conversioneff1}) the conversion efficiency
becomes

\begin{eqnarray}\label{eficw}
    \eta_{cw} &=&\frac{2^{2}3^{2}\hbar^{2} c^{3}n_p^{3}\gamma^{2}L^{2}}{\pi^2\omega_{p}}\nonumber\\
    &\times&\int\!\! d\omega_r\!\! \int\!\! d\omega_s\, h(\omega_r,\omega_s,\omega_{p}-\omega_r-\omega_s)\nonumber\\
    &\times&\mbox{sinc}^{2}\left[\frac{L}{2}\Delta
    k_{cw}(\omega_r,\omega_s)\right].
\end{eqnarray}

In Eqns.~(\ref{Ncw}) and (\ref{eficw}) $\omega_p$ is the frequency
of the monochromatic-pump. These equations have been written in
terms of the phase mismatch $\Delta k_{cw}(\omega_r,\omega_s)$ [see
Eq.~(\ref{Eq:PM})] defined as

\begin{eqnarray}
    \Delta
    k_{cw}(\omega_r,\omega_s)&=&k(\omega_p)-k(\omega_r)-k(\omega_s)\nonumber\\&-&k(\omega_{p}-\omega_r-\omega_s)+\Phi_{NL},
\end{eqnarray}

\noindent and the function
$h(\omega_r,\omega_s,\omega_{p}-\omega_r-\omega_s)$ defined
as

\begin{equation} \label{Eq:hfunct}
    h(\omega_r,\omega_s,\omega_i) \equiv \frac{k_r'\omega_r}{n_r^2}\frac{k_s'\omega_s}{n_s^2}
    \frac{k_i'\omega_i}{n_i^2}.
\end{equation}

In order to gain a better understanding of the TOSPDC process, we
show in the next subsection that it is possible to obtain emitted
flux expressions in closed analytic form under certain
approximations.

\subsection{Non-degenerate emission frequencies: Closed analytic expressions}\label{secAnali}

In order to obtain a closed analytic expression for the emitted
flux, we start by considering that the function
$h(\omega_r,\omega_s,\omega_i)$, contained by the integrand in
Eq.~(\ref{Eq:NumFot}), varies only slowly with the generation
frequencies, within a sufficiently narrow spectral
region of interest.  Thus, in what follows we approximate this
function to be constant; specifically, we evaluate the function
$h(\omega_r,\omega_s,\omega_i)$~[see Eq.~(\ref{Eq:hfunct})] at the
frequencies $\omega_{\mu 0}$ (where $\mu=r,s,i$), for which perfect
phasematching is attained.

In addition, in order to solve the triple frequency integral in
Eq.~(\ref{Eq:NumFot}) we resort to a linear approximation of the
phasemismatch. Within this approximation, it can be shown that the
product $L\Delta k$ in the phase matching function [see
Eq.~(\ref{fphmat})] can be expressed as

\begin{equation} \label{Eq:pmlin}
     L\Delta k_{lin}=\tau_r\nu_r+\tau_s\nu_s+\tau_i\nu_i,
\end{equation}

\noindent written in terms of the frequency detunings
$\nu_{\mu}=\omega_{\mu}-\omega_{\mu 0}$.  In Eq.~(\ref{Eq:pmlin}),
we have assumed that the constant term of the Taylor expansion
vanishes, i.e. that phasematching is attained at the central pump
and generation frequencies $\omega_{\mu 0}$ (with $\mu=p,r,s,i$).
Parameters $\tau_{\mu}$ represent group velocity mismatch
coefficients between the pump and each of the emitted modes, and are
given by $\tau_{\mu}=L(k'_{p0}-k'_{\mu 0})$, where
$\mu=r,s,i$.

We also assume that before reaching the detectors, the TOSPDC
photons (in each of three modes) are transmitted through gaussian spectral
filters of bandwidth  $\sigma_{f\mu}$, represented by the function
$f_{fil}=\exp(-\nu_{\mu}^{2}/\sigma_{f\mu}^{2})$ (with $\mu=r,s,i$). The resulting
filtered joint spectral amplitude function, assuming that all three filters have the same bandwidth $\sigma_f$, is given by

\begin{eqnarray}\label{jsafil}
    f_{fil}(\nu_r,\nu_s,\nu_i)&=&f(\nu_r,\nu_i,\nu_s)\exp\left[-\frac{\nu_r^{2}+\nu_s^{2}+\nu_i^{2}}{\sigma_{f}^{2}}\right].\nonumber\\
\end{eqnarray}

Then, by replacing Eqns.~(\ref{Eq:hfunct}), (\ref{Eq:pmlin}), and
(\ref{jsafil}) into Eq.~(\ref{Eq:NumFot}) it can be shown that the
number of photon triplets emitted per second is given by

\begin{eqnarray}\label{Napprox}
    N &=& \frac{3^{2}\hbar c^{3}n_p^{3}}{(2\pi)\omega_{p0}^2}
    \frac{L^{2}\gamma^{2} p \sigma_f^{3}}{(\sigma^{2}+3\sigma_f^{2})^{1/2}}h(\omega_{r0},\omega_{s0},\omega_{i0})\nonumber\\
    &\times&\frac{1}{\Phi}\{2\sqrt{\pi\Phi}\mathrm{erf}[2\sqrt{\Phi}]+\exp(-4\Phi)-1\},
\end{eqnarray}

\noindent where $\text{erf}(.)$ denotes the error function and
$\Phi$ is given by

\begin{eqnarray}\label{parPhi}
    \Phi&=&\frac{\sigma_f^{2}}{32(\sigma^{2}+3\sigma_f^{2})}\bigg[(\sigma^{2}+2\sigma_f^{2})(\tau_r^{2}+\tau_s^{2}+\tau_i^{2})\nonumber\\
    &-&2\sigma_f^{2}(\tau_r\tau_s+\tau_r\tau_i+\tau_s\tau_i)\bigg].
\end{eqnarray}

We will concentrate our further discussion on the specific case
where the filter bandwidth $\sigma_f$ is much greater than the pump
bandwidth $\sigma$.  This scenario is realistic for a pump in the
form of a picosecond-duration pulse train, as will be studied in the
context of a specific example in Sec.~\ref{designs}. In this case,
$\Phi$ reduces to $\Phi=(L/L_0)^2$, in terms of a characteristic
length $L_0$ given by

\begin{equation}
\label{lo} L_0= \frac{\sqrt{48}}{\sigma_f}
\!\frac{1}{\sqrt{{k_{r0}'}^2+{k_{s0}'}^2+{k_{i0}'}^2-k_{r0}'
k_{s0}'-k_{r0}' k_{i0}'-k_{s0}' k_{i0}'}}.
\end{equation}

Let us note that for  $\sigma_f \gg \sigma$, Eq.~(\ref{Napprox})
diverges for frequency-degenerate TOSDPC for which
$k'_{r0}=k'_{s0}=k'_{i0}$, due to the $1/\Phi$ dependence.   Indeed,
the linear approximation of the phasemismatch employed here fails
for frequency-degenerate TOSPDC, unless the emission modes are
strongly filtered (i.e. $\sigma_f \ll \sigma$).  While the PM
function $\phi(\omega_r,\omega_s,\omega_i)$ has a curvature in the
emission frequencies space $\{\omega_r,\omega_s,\omega_i\}$ which
limits the overlap with the PSA function
$\alpha(\omega_r+\omega_s+\omega_i)$, the linearly-approximated PM
function has the same orientation as the PSA function, which leads
to the unphysical situation of an infinite emission bandwidth, in
turn leading to the above-mentioned divergence.   Thus, we restrict
the use of the expression in closed analytic form for the emitted
flux to the case of frequency non-degenerate TOSPDC.  As we will
study in Sec.~\ref{designs}, our flux expression in closed analytic
form for the non-degenerate case leads to excellent agreement with a
numerical calculation which does not resort to approximations.

Let us now consider two different limits of Eq.~(\ref{Napprox}).
Note that for a sufficiently large $\Phi$ value,
$[(2\sqrt{\pi\Phi}\mathrm{erf}[2\sqrt{\Phi}]+\exp(-4\Phi)-1)]/\Phi$
becomes $2 \sqrt{\pi / \Phi}$.  Let us denote by $\phi$ a $\Phi$
value so that for $\Phi \gtrsim \phi$, this limit has been reached.
For example, for $\Phi > 100$ which corresponds to $L> 10 L_0$ the
above function approaches this limit within $<3\%$.

Thus, for $L\gtrsim \sqrt{\phi} L_0$, the number of photon triplets emitted per pump pulse can be well approximated by

\begin{align}\label{NanalargeL}
N&=\frac{6^2 \hbar c^3k_{r0}' k_{s0}' k_{i0}'}{\sqrt{\pi}\sqrt{{k_{r0}'}^2+{k_{s0}'}^2+{k_{i0}'}^2-k_{r0}' k_{s0}'-k_{r0}' k_{i0}'-k_{s0}' k_{i0}'}} \nonumber \\
&\times \frac{ \omega_{r0} \omega_{s0} \omega_{i0}}{\omega_{p0}^2}
\frac{n_{p0}^3}{n_{r0}^2 n_{s0}^2 n_{i0}^2} \gamma^2 L p \sigma_f.
\end{align}

Conversely, for $L \lesssim \sqrt{\phi} L_0$, the number of photon triplets emitted per pump pulse becomes

\begin{align}
N&=\frac{18 \hbar c^3}{\sqrt{3} \pi} k_{r0}' k_{s0}' k_{i0}'
 \frac{ \omega_{r0} \omega_{s0} \omega_{i0}}{\omega_{p0}^2} \frac{n_{p0}^3}{n_{r0}^2 n_{s0}^2 n_{i0}^2} \gamma^2 L^2 p \sigma_f^2.
\end{align}

Thus, while for a short fiber (compared to $ \sqrt {\phi} L_0$) the
flux vs fiber length is quadratic, for longer fiber lengths this
dependence becomes linear.  Note that $L_0$ represents a measure of
the wavepacket length for each of the three emitted modes.  Thus,
the quadratic dependence appears for fibers which have a length
similar or shorter as compared to the emitted wavepacket length. For
most situations of interest, $L_0$ is a small quantity; indeed, as
will be the case for the particular example studied in
Sec.~\ref{Long}, the flux dependence with fiber length can be
regarded as linear, as given by Eq.~(\ref{NanalargeL}).

This analysis serves to clarify the dependence of the emitted flux
on all experimental parameters of interest, in the case of
non-degenerate TOSPDC.  The emitted flux is linear with respect to
$L$, constant with respect to $\sigma$ and linear with respect to
$p$.   Although the analytic expressions which we have obtained are
not valid for frequency-degenerate TOSPDC, our numerical results
(see Sec.~\ref{flux_design}) indicate a qualitatively identical
dependence of the emitted flux vs these experimental parameters.

The observed behavior for TOSDPC is different from that observed for
the SFWM process, for which the emitted flux is linear in $\sigma$
\cite{garay10}. This means that shorter pump pulses do not lead to
higher rates of emission for TOSPDC, as is the case for SFWM. Note
that the manner in which the emitted flux depends on various
experimental parameters is essentially identical to the behavior
observed for spontaneous parametric downconversion in crystals with
a second-order nonlinearity.

\section{TOSPDC phasematching proposal}\label{technique}

A crucial aspect in the design of a photon-triplet TOSPDC source is
the need for phasematching between the four participating fields.
Specifically, this translates into the condition $\Delta
k(\omega_{r0},\omega_{s0},\omega_{i0})=0$ [see Eq.~(\ref{Eq:PM})],
for a given central pump frequency.

In general, it is not trivial to fulfill phasematching for TOSPDC
due to the large spectral separation between the pump and the
emitted photons; for the frequency-degenerate case, pump photons at
frequency $3 \omega$ are annihilated in order to generate photon
triplets at $\omega$.  For most common materials including fused
silica, $k(3\omega)$ is considerably larger than $3k(\omega)$, while
these two quantities must be equal for the TOSPDC process operated
in the low pump-power limit to be phasematched.  We have proposed
(see Ref.~\cite{corona11}) a multi-modal phasematching strategy, in
which the pump mode propagates in a different fiber mode compared to
the generated TOSPDC photons.  Note that similar strategies have
been exploited for third-harmonic
generation~\cite{kolevatova03,grubsky07}. Specifically, we assume
that the pump mode propagates in the first excited mode
($\mathrm{HE}_{12}$), while the signal-1, signal-2, and idler
photons propagate in the fundamental mode ($\mathrm{HE}_{11}$) of
the fiber~\cite{corona11}.  This technique permits phasematching  at
the cost of limiting the attainable mode overlap between the pump
and the TOSDPC modes.  Furthermore, the fact that the pump must
propagate in the $\mathrm{HE}_{12}$ mode for our phasematching
strategy limits the power than can be coupled from, say, a
Gaussian-transverse-distributed pump mode in free space; this will
tend to limit the attainable source brightness.

We focus our attention on thin fused silica fibers guided by air,
i.e. where the core is a narrow fused silica cylinder, and the
cladding is the air surrounding this core. The combination of a
small fiber diameter and a large core-cladding index of refraction
contrast leads to a strong waveguide contribution to the overall
dispersion experienced by the propagating fields which can enhance
nonlinear optical effects, including TOSDPC. Note that similar
results could be obtained with photonic crystal fibers involving a
large air-filling fraction in the cladding.  Note also that the
non-ideal TOSDPC-pump overlap observed for our multi-modal
phasematching approach can to some degree be compensated by the
small transverse mode area, which tends to enhance the nonlinearity
$\gamma$.

In general, for a particular set of desired pump and TOSPDC
frequencies, we find that a specific fiber radius can exist, to be
referred to as phasematching radius, for which phasematching is
attained.  For optical frequencies of interest, phasematching radii
tend to be in the sub-micrometer core diameter range.   It is worth
mentioning that such fiber radii can be obtained through current
fiber taper technology (e.g. see
Refs.~\cite{tong03,lsaval04,brambilla10}).

\begin{figure}[ht]
\centering\includegraphics[width=6 cm]{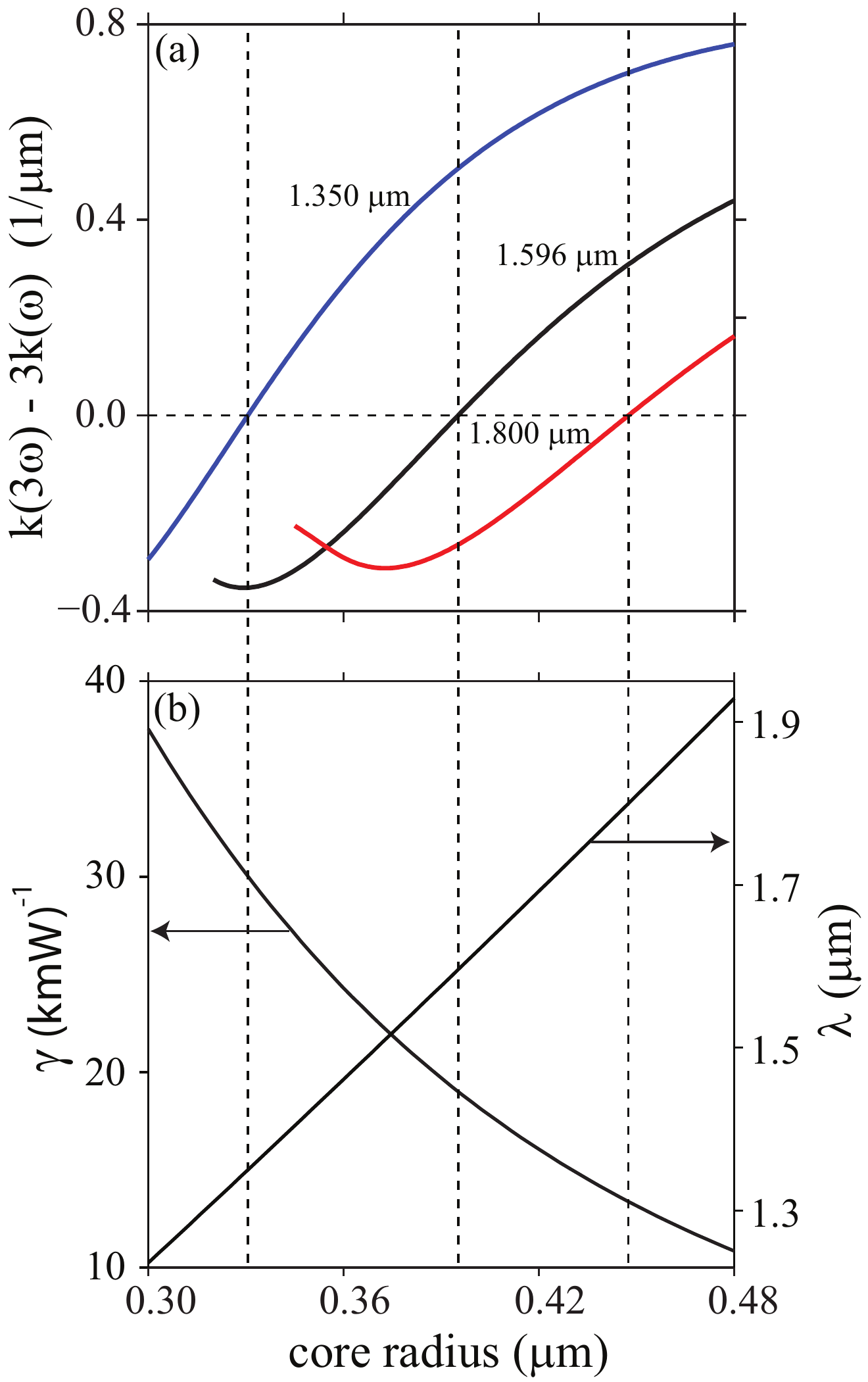}\caption{(color online) (a)
Frequency-degenerate phase-matching for TOSPDC at $1.350\mu$m (blue
line), $1.596\mu$m (black line), and $1.800\mu$m (red line). (b)
Degenerate TOSPDC wavelength, and non-linear coefficient $\gamma$ vs
phasematching radius.} \label{Fig:PM1}
\end{figure}

As an illustration, in Fig.~\ref{Fig:PM1}(a) we plot as a function
of the core radius the phasemismatch  $k_p(3 \omega)-3
k_{rsi}(\omega)$,  for three different choices of the emitted
frequency: $\omega=2\pi c/1.350\mu \mbox{m}$, $\omega=2\pi
c/1.596\mu \mbox{m}$, and $\omega=2\pi c/1.800\mu \mbox{m}$, where
functions $k_p(\omega)$ and $k_{rsi}(\omega)$ are evaluated for the
HE$_{12}$ and HE$_{11}$ modes, respectively. As is clear from this
figure, the low-pump-power phase-matching condition
$k_p(3\omega)=3k_{rsi}(\omega)$ is fulfilled for a specific core
radius for each of the considered $\omega$ values: $r=0.331\mu$m,
$r=0.395\mu$m, and $r=0.448\mu$m, respectively. In
Fig.~\ref{Fig:PM1}(b) we show the general trend for the degenerate
TOSPDC frequency (expressed in terms of wavelength) vs phasematching
radius, where the dotted vertical lines denote the specific
frequencies considered in Fig.~\ref{Fig:PM1}(a). From this figure we
can see that core radii in the range $300 -480$nm are required for
degenerate TOSPDC wavelengths within the range $1.24-1.93\mu$m. Note
that while we have concentrated here on frequency-degenerate TOSPDC,
this technique can also be extended to the frequency non-degenerate
case.

In Fig.~\ref{Fig:PM1}(b) we also show the nonlinear coefficient
$\gamma$ [see Eq.~(\ref{Eq:gam3})] on the phasematching radius,
for the case of frequency-degenerate TOSPDC. Note
that decreasing the core radius leads to an increase in the
phasematched degenerate TOSPDC frequency, and likewise to an increase
in the nonlinearity $\gamma$.

\section{Design considerations for photon-triplet sources}\label{considerations}

In this section we focus on the general considerations that should
be taken into account in designing a TOSPDC photon-triplet source.
Of particular interest is the choice of pump and TOSPDC frequencies.
For the type of fiber considered in this paper, i.e. constituted by
a fused silica core and where the cladding is the air surrounding
this core, the generation frequencies depend on two parameters: the
fiber radius and the pump frequency. Note that while the
phasemismatch has a pump-power dependence [see Eq.~(\ref{Eq:PM})],
the overall pump-power dependence of emission frequencies tends to
be negligible for pump-power levels regarded as typical.

\begin{figure}[ht]
\centering \includegraphics[width=8 cm]{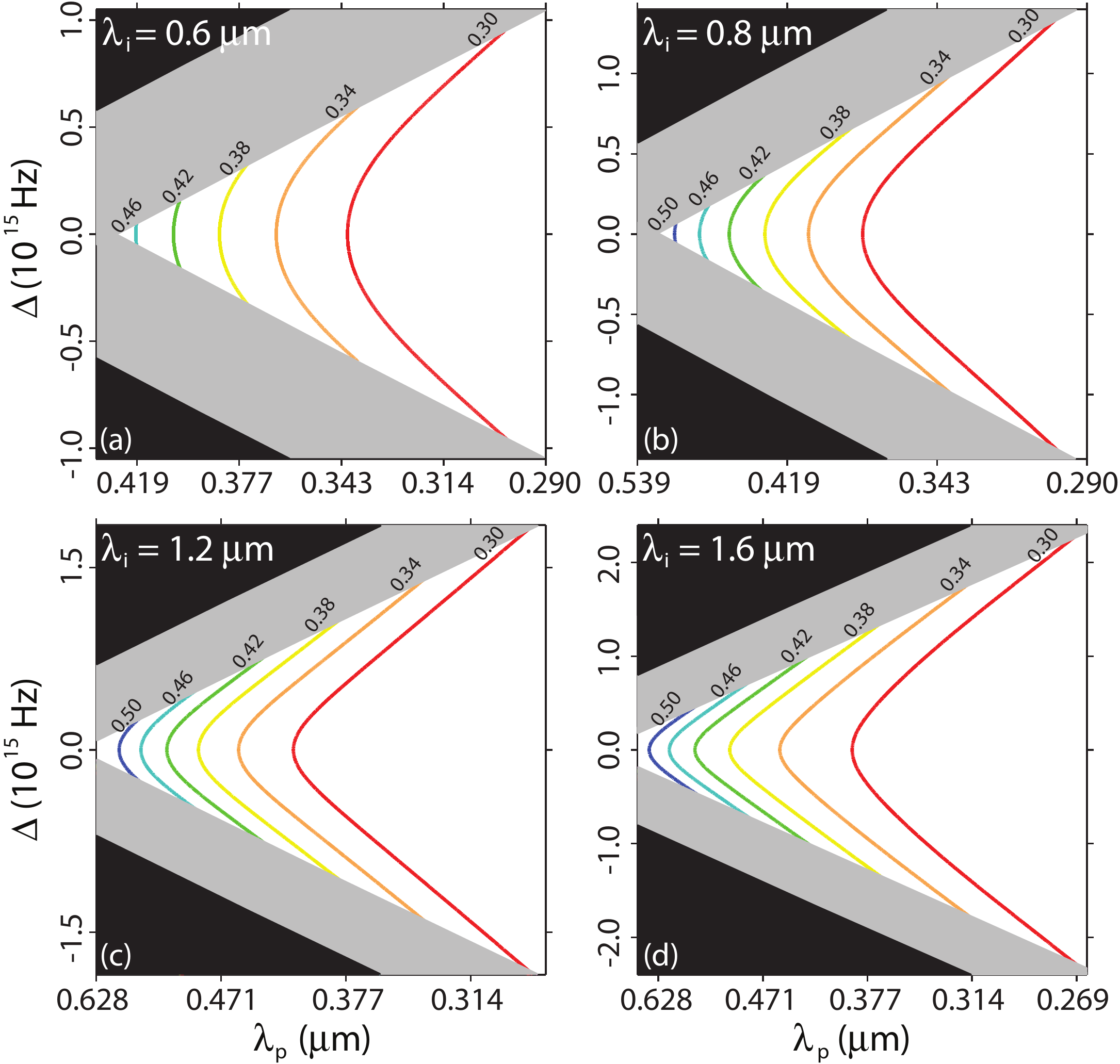}\caption{(color
online)  Phasematched emission frequencies plotted as a function
of the pump frequency, for different fiber radii, and assuming
the following idler wavelengths, kept constant for each of the four panels:
(a) $\lambda_i=0.6\mu$m, (b)
$\lambda_i=0.8\mu$m, (c) $\lambda_i=1.2\mu$m, and (d)
$\lambda_i=1.6\mu$m.} \label{Fig:PM_diagrams}
\end{figure}

In Fig.~\ref{Fig:PM_diagrams}, we present a characterization of the
emission frequencies as a function of the core radius and the pump
frequency.  Each of the four panels shown [(a) through (d)]
corresponds to a fixed value of the idler frequency $\omega_i$.  In
particular, we have chosen the following values of $\omega_i$: (a)
$\omega_i=2\pi c/0.6\mu\mbox{m}$, (b) $\omega_i=2\pi
c/0.8\mu\mbox{m}$, (c) $\omega_i=2\pi c/1.2\mu\mbox{m}$, and (d)
$\omega_i=2\pi c/1.6 \mu\mbox{m}$. In each panel, we have plotted
the phasematched signal-1(r) and signal-2(s) emission frequencies
expressed as the frequency detunings
$\Delta_r=\omega_r-(\omega_p-\omega_i)/2$ and
$\Delta_s=\omega_s-(\omega_p-\omega_i)/2$, respectively, as a
function of the pump frequency $\omega_p$; note that energy
conservation implies that $\Delta_r=-\Delta_s$, and we define
$\Delta \equiv \Delta_r$.    Specifically, each curve gives
combinations of pump, signal-1 and signal-2 frequencies yielding
perfect phasematching, i.e. $k_p-k_r-k_s-k_i=0$ (where we have
neglected the nonlinear phase term $\Phi_{NL}$). Different curves in
a given panel were calculated for a choice of different values of
the core radius (within the range $r=0.3-0.5\mu $m). In all four
panels, gray-shaded areas represent regions of the
$\{\omega_p,\Delta_{r,s}\}$ space for which $\omega_r$ and/or
$\omega_s$ lie outside of the range of validity of the dispersion
relation used for fused silica. Non-physical zones for which
$\omega_r$ and/or $\omega_s$ would have to be negative in order to
satisfy energy conservation are shaded in black.

Form these curves, it can be appreciated that for each $\omega_i$, there is
a continuum of core radii for which phasematching occurs. For each $\omega_i$,
while the core radius can
be reduced without limit and still obtain perfect phasematching (within the spectral
window considered here), a
maximum core radius exists, above which phasematching
is no longer possible.  Indeed, as the core radius is increased, the spread of
$\Delta_r$ and $\Delta_s$ values is reduced until it reaches the single value
$\Delta_r=\Delta_s=0$.  Likewise, note that for a fixed core radius, the spread of
$\Delta_r$ and $\Delta_s$ values shrinks for higher values of $\omega_i$.
Note that the
vertex of the phase-matching contours indicates the emission of triplets
for which $\Delta_{r,s}=0$, or equivalently, $\omega_r=\omega_s$. Note that
for a particular $\omega_i$ value, there is a single core radius for which this
vertex corresponds to the frequency-degenerate emission, i.e. with
$\omega_r=\omega_s=\omega_i$.

\begin{figure}[ht]
\centering\includegraphics[width=7 cm]{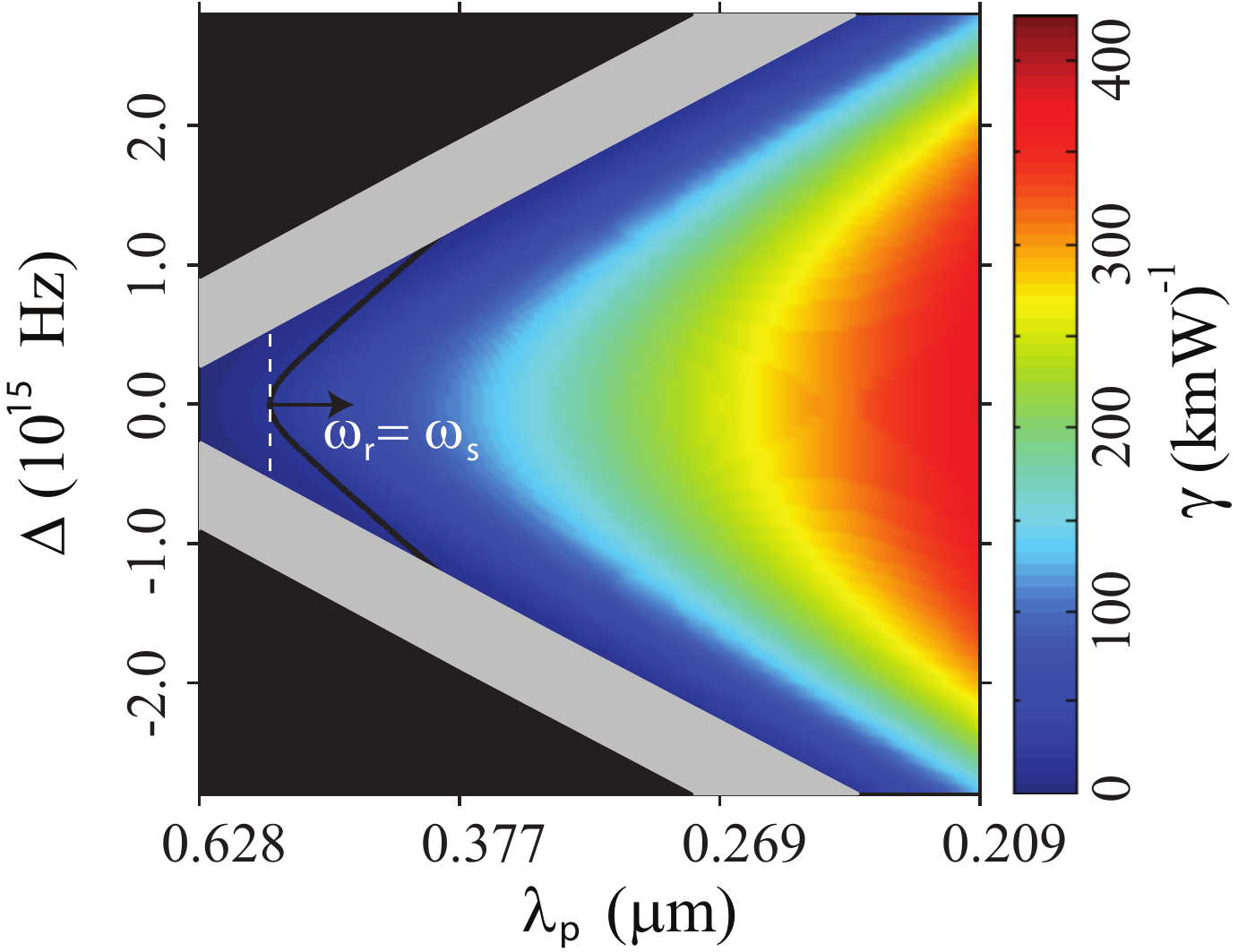}\caption{(color online)
Nonlinear coefficient $\gamma$ as a function of $\omega_p$ and
$\Delta$, for $r=0.395\mu$m and $\omega_i=2\pi
c/1.596\mu\mbox{m}$.  The black-solid line represents frequency combinations
leading to perfect phasematching.}
\label{Fig:Gamma}
\end{figure}

Experimental constraints such as available pump frequencies,
spectral windows of single-photon detectors, and attainable fiber
radii may in principle be used together with the curves in
Fig.~\ref{Fig:PM_diagrams} in order to determine the required source
parameters.  An important  aspect to consider is the nonlinearity
$\gamma$ [given by Eq.~(\ref{Eq:gam3})], which of course has an
impact on the source brightness; indeed, from
Eq.~(\ref{Eq:conversioneff1}), it is clear that the conversion
efficiency scales quadratically with $\gamma$.  In general, $\gamma$
is determined by the core radius $r$, as well as by the pump and
emission frequencies. In Fig.~\ref{Fig:Gamma} we present for a fixed
radius ($r=0.395\mu$m) and a fixed idler frequency ($\omega_i=2 \pi
c/1.596\mu$m) a plot of $\gamma$ vs $\Delta$ and $\omega_p$. In this
figure the value of $\gamma$ for each $(\omega_p,\Delta)$ point is
indicated by the colored background, regardless of whether or not
phase matching is achieved at that point. It can be seen from this
figure that significantly higher  values of $\gamma$ are obtained
for large pump frequencies (lying in the ultraviolet region of the
optical spectrum), and for $\Delta_{r,s}\rightarrow 0$ i.e.,
$\omega_r\rightarrow \omega_s$.  The black line in
Fig.~\ref{Fig:Gamma} represents the contour formed by phasematched
frequencies.   Thus, unfortunately, the highest $\gamma$ values are
inaccessible because they occur for unphasematched frequency
combinations.

\section{Specific TOSPDC photon-triplet source
designs}\label{designs}

From the discussion in Sec.~\ref{considerations}, it is clear that
in order to optimize the nonlinearity [see Figs.~\ref{Fig:PM1}(b)
and \ref{Fig:Gamma}] small core radii and large pump frequencies are
required.  While this might suggest the use of an ultraviolet pump,
in this paper we avoid the use of non-standard fiber-transmission
frequencies.  Thus, we propose source designs for which the pump
frequency is in the region of $0.532\mu$m, which for
frequency-degenerate TOSPDC results in photon triplets centered
around $1.596\mu$m.

\begin{figure}[ht]
\centering\includegraphics[width=8 cm]{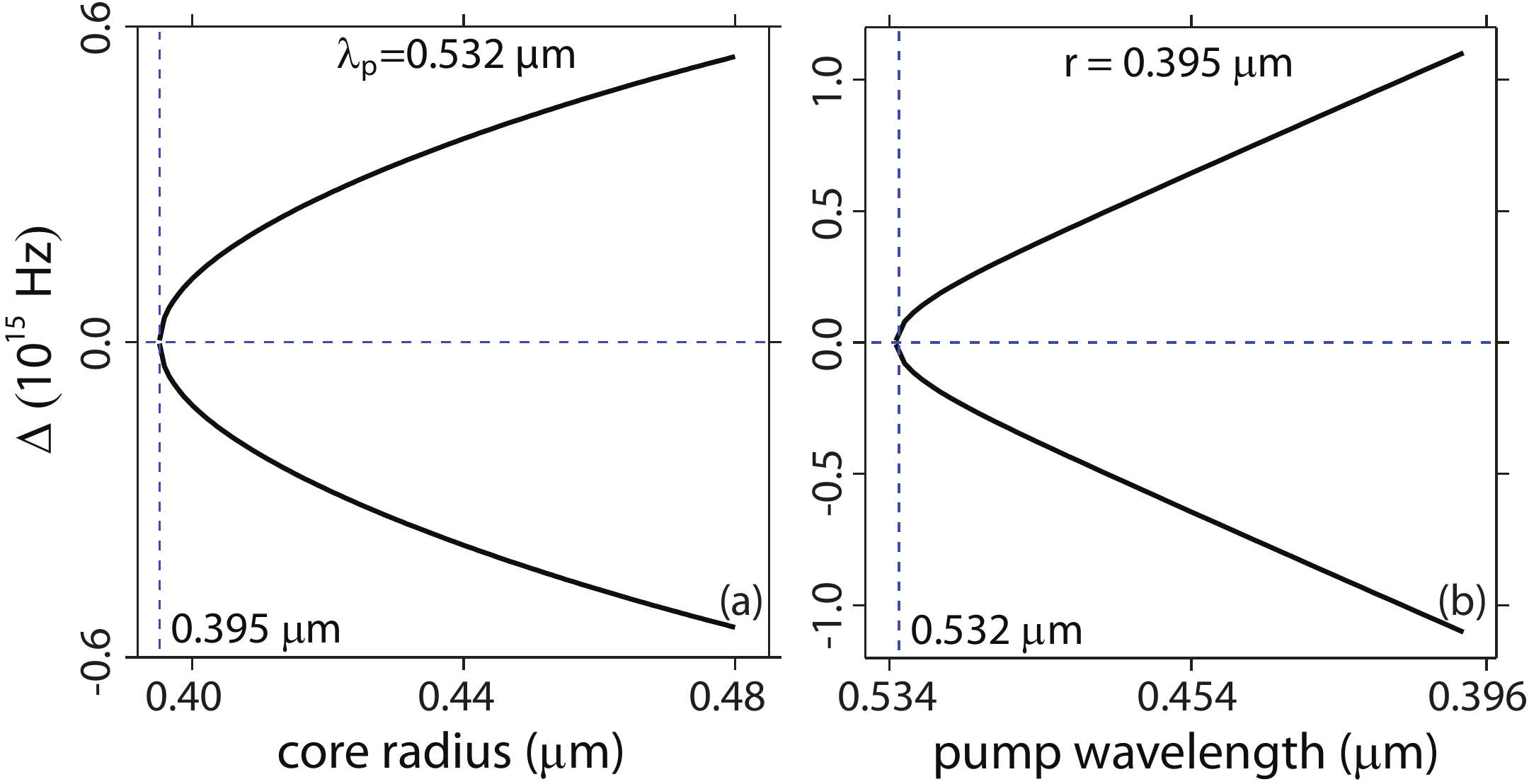}\caption{(color online) (a) Phasematched emission frequencies
as a function of the core radius, for a fixed pump wavelength ($\lambda_p=0.532\mu\mbox{m}$).   (b) Phasematched emission frequencies
as a function of the pump frequency, for a fixed fiber radius ($r=0.395\mu$m).}
\label{Fig:Gamma2}
\end{figure}

Let us initially assume that the pump frequency is given by
$\omega_p=2\pi c/0.532\mu\mbox{m}$ and let us fix the idler
frequency to $\omega_i=\omega_p/3=2\pi c/1.596\mu\mbox{m}$.
Fig.~\ref{Fig:Gamma2}(a) shows the resulting emission frequencies,
displayed in terms of the detuning variable $\Delta$, plotted vs the
core radius $r$.   From this figure, it is clear that there is a
specific core radius ($r=0.395\mu m$) for which the emission
frequencies are characterized by $\Delta=0$, which in this case
implies $\omega_r=\omega_s=\omega_i$, i.e. for which the TOSPDC
process is frequency degenerate.   Note from the figure that
decreasing the core radius from the value $r=0.395\mu m$ leads to
the suppression of phasematching.  Likewise, note that increasing
the core radius from this value, leads to $\Delta \neq 0$, so that
$\omega_r=\omega_p/3+\Delta$ and $\omega_s=\omega_p/3-\Delta$.  In
other words, the three emission frequencies become distinct, leading
to frequency non-degenerate TOSPDC.   Thus, with a fixed pump
frequency the core radius is a useful experimental parameter for the
control of the degree of frequency non-degeneracy.

A similar behavior is observed by making the fiber radius, instead
of $\omega_p$, constant (to a value of $r=0.395\mu$m), while varying
$\omega_p$. Fig.~\ref{Fig:Gamma2}(b) shows the resulting emission
frequencies, displayed in terms of the detuning variable $\Delta$,
plotted vs the pump frequency $\omega_p$.   From this figure, it is
clear that for a pump frequency of $\omega_p=2 \pi c/0.532\mu$m, the
resulting emission frequencies are characterized by $\Delta=0$,
which in this case implies frequency degenerate TOSPDC with
$\omega_r=\omega_s=\omega_i$.   Note from the figure that decreasing
$\omega_p$ from  a value of $\omega_p=2 \pi c/0.532\mu$m leads to
the suppression of phasematching.  Likewise, note that increasing
$\omega_p$ from  this value, leads to $\Delta \neq 0$, so that
$\omega_r=(\omega_p-\omega_i)/2+\Delta$ and
$\omega_s=(\omega_p-\omega_i)/2-\Delta$. In other words, the three
emission frequencies become distinct, leading to frequency
non-degenerate TOSPDC.   Thus, with a fixed core radius, the pump
frequency  is a useful experimental parameter for the control of the
degree of frequency non-degeneracy.

Throughout the rest of this paper, we will consider two source designs, both
based on a fiber of radius $r=0.395\mu$m and length $L=10$cm.

\begin{itemize}

\item Frequency degenerate source, with $\omega_p=2 \pi c/0.532\mu$m and with emission modes centered at: $\omega_r=\omega_s=\omega_i=2 \pi c/1.596\mu$m.

\item Frequency non-degenerate source, with $\omega_p=2 \pi c/0.531\mu$m and with emission modes centered at: $\omega_i=2 \pi c/1.596 \mu$m, $\omega_r=2 \pi c/1.529\mu$m, and $\omega_i=2 \pi c/1.659\mu$m.   As will be discussed in the next subsection, in order to guarantee that the emission modes are spectrally distinct, frequency
    filters should be used.

\end{itemize}

In what follows, we show plots of the joint spectral intensity
function for these TOSPDC photon-triplet source designs.

\subsection{Joint spectral intensity of the proposed TOSPDC sources}

In this section, we present representations of the TOSPDC photon-triplet
state for the source designs proposed above.  Such plots
are useful in order to visualize the spectral correlations which underlie
the existence of  entanglement in the photon triplets.

When plotted in the generation frequencies space $\{\omega_s,\omega_r,\omega_i\}$
for typical experimental parameters, the joint spectrum of the frequency-degenerate TOSPDC
state is akin to a
``membrane'' of narrow width along the direction $\omega_s+\omega_i+\omega_r$, and
much larger widths along the two perpendicular directions.  In the limiting case
of a monochromatic pump, this membrane becomes infinitely narrow, leading
to spectrally anti-correlated photon triplets, with the sum of the three generation
frequencies $\omega_s+\omega_r+\omega_i$ equal to a constant value, $\omega_p$.

In this paper we have used two different approaches for the
visualization of the JSI. On the one hand, it is useful to
re-express the joint amplitude function [see Eq.~(\ref{Eq:FF})] in
terms of frequency variables which are chosen in accordance to the
symmetry exhibited by the quantum state.   Thus, we use variables
$\{\nu_A,\nu_B,\nu_{+}\}$ obtained by an appropriate rotation of the
frequency detuning axes $\{\nu_r,\nu_s,\nu_i\}$ so that the new
$\nu_A$ and $\nu_B$ axes are tangent to the perfect phasematching
surface contour (which, again, is akin to a tilted membrane, in this
case with vanishing width), and so that the $\nu_{+}$ axis is normal
to this surface contour.   The transformation between these two sets
of frequency variables is

\begin{align}
\nu_{+}&=\frac{1}{\sqrt{3}}(\omega_r+\omega_s+\omega_i-3\omega_0)  \nonumber \\
\nu_A&=\frac{1}{2}\left(1-\frac{1}{\sqrt{3}}\right) \omega_r + \frac{1}{2}\left(-1-\frac{1}{\sqrt{3}}\right) \omega_s+\frac{1}{\sqrt{3}} \omega_i \nonumber \\
\nu_B&=\frac{1}{2}\left(1+\frac{1}{\sqrt{3}}\right) \omega_r +
\frac{1}{2}\left(-1+\frac{1}{\sqrt{3}}\right)
\omega_s-\frac{1}{\sqrt{3}} \omega_i.
\end{align}

\begin{figure}[ht]
\centering\includegraphics[width=7 cm]{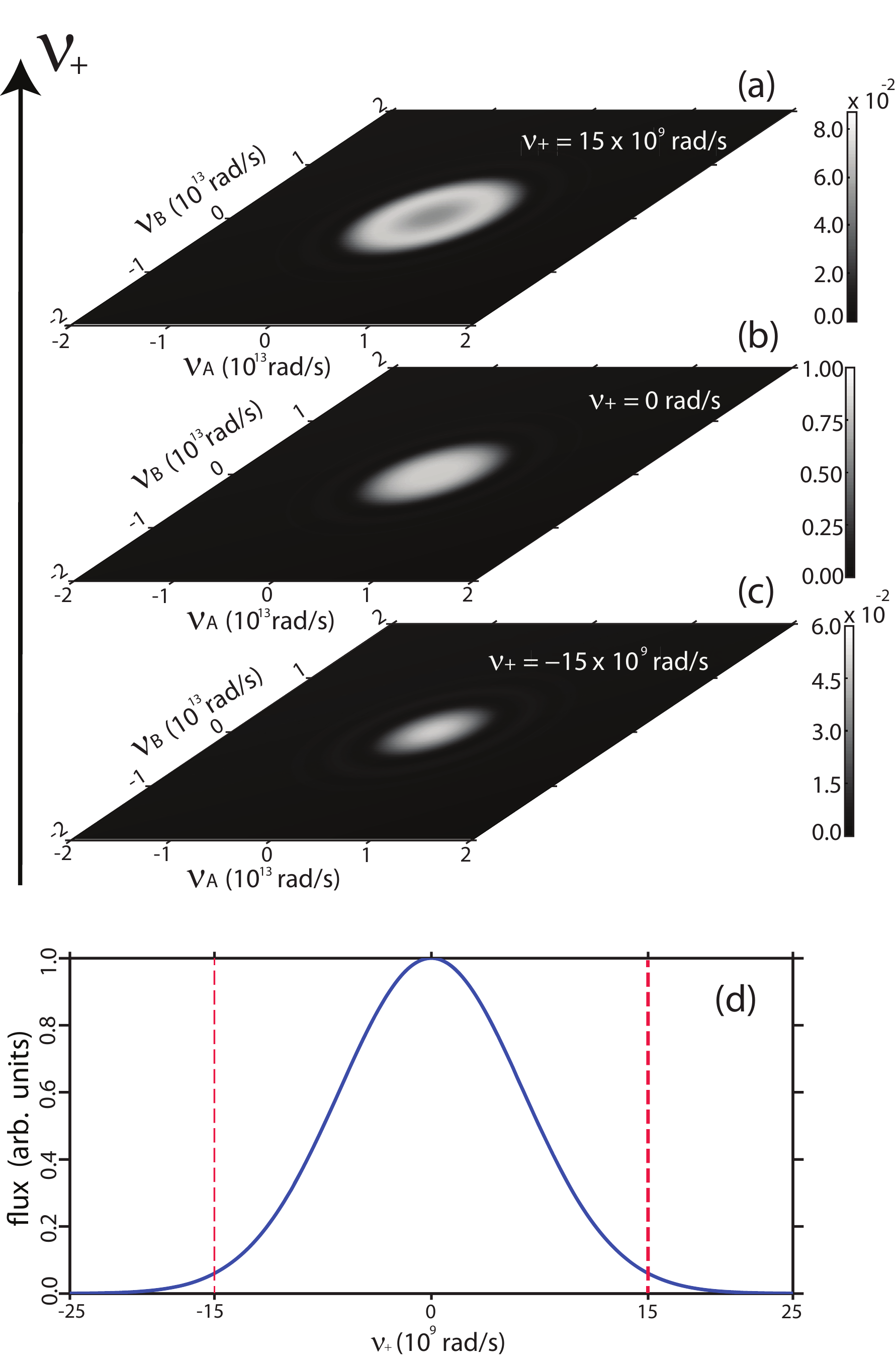}{0.4}\caption{(color online)
Representation of the JSI for our frequency-degenerate TOSPDC source
design, plotted as a function of the frequency variables $\nu_A$ and
$\nu_B$ for the following fixed values of $\nu_{+}$:
$\nu_+=15\times10^{9}\,\mbox{rad/s}$ (a) , $\nu_+=0$ (b) and
$\nu_+=-15\times10^{9}\,\mbox{rad/s}$ (c). (d) JSI plotted as a
function of $nu_{+}$, for $\nu_A=\nu_B=0$.}\label{figest_rotado}
\end{figure}

We may write down a version of the joint amplitude function in terms
of these new frequency variables, $f'(\nu_{A},\nu_{B},\nu_{+})$, by
expressing each of the original variables in terms of the new ones.
In Fig.~\ref{figest_rotado}, panels (a) through (c), we have plotted
the JSI function $|f'(\nu_{A},\nu_{B},\nu_{+})|^2$ resulting from
making $\nu_{+}$ constant to one of three different values: $-15$GHz
(panel c), $0$ (panel b) and $15$GHz (panel a), for the following
choice of parameters: $L=10$cm, $\omega_p=2 \pi c/0.532nm$, and
$\sigma=23.5$GHz (this corresponds to one frequency-degenerate
TOSPDC source design). These three plots can be thought of as
distinct `slices' of the three-dimensional JSI at different
$\nu_{+}$ values. Panel (d) represents a plot of the JSI function
$|f'(0,0,\nu_{+})|^2$, i.e. the choice of variables which are left
constant and those that are allowed to vary are reversed. Thus,
while the plot in Fig.~\ref{figest_rotado}(b) gives the relatively
large transverse extension of the ``membrane'' referred to in the
previous paragraph, Fig.~\ref{figest_rotado}(d) gives the much
smaller longitudinal width of the ``membrane''.  From a graphical
analysis of panels (a)-(c), it is clear that making $\nu_{+}$
negative leads to a suppression of phasematching, while making
$\nu_{+}$ positive leads to a ring structure, implying that the
``membrane'' referred to above is actually curved.   Note that the
width of the curve in Fig.~\ref{figest_rotado}(d) can approach zero
either in the case of a very narrow pump bandwidth or in the limit
of a very long fiber.

\begin{figure}[ht]
\centering\includegraphics[width=8 cm]{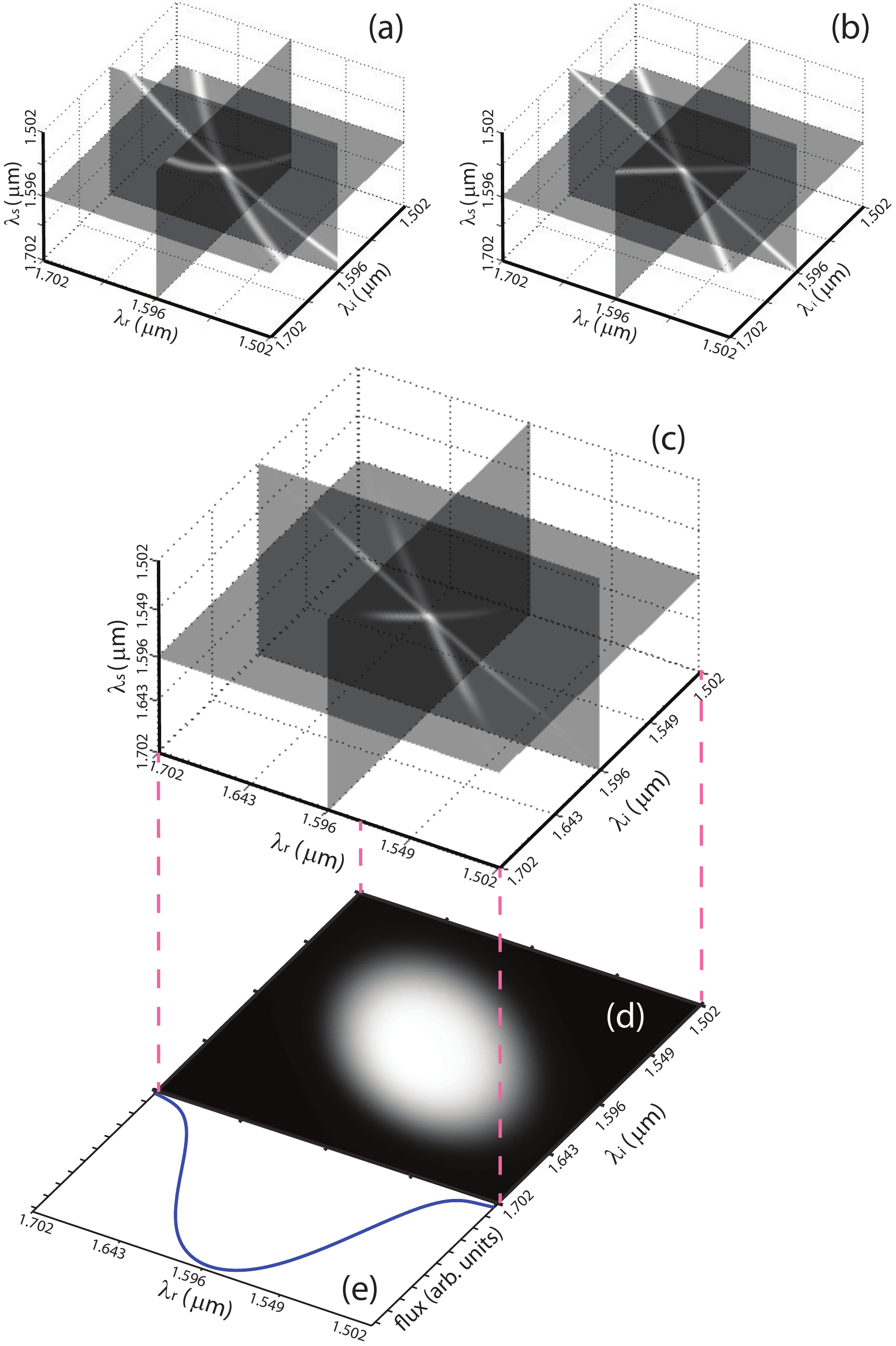}\caption{(color online) Plotted as
function of the three emitted frequencies
$\{\omega_r,\omega_s,\omega_i\}$:  (a) Phase matching function
$|\phi(\omega_r,\omega_s,\omega_i)|^2$, (b) pump spectral amplitude
$|\alpha(\omega_r+\omega_s+\omega_i)|^2$, and (c) JSI. (d)
Two-photon spectrum $I_2(\omega_r,\omega_s)$. (e) Single-photon
spectrum $I_1(\omega_r)$. Panels (a)-(c) are similar to a figure
from Ref.~\cite{corona11}.}\label{fig3Dstate}
\end{figure}

It is also useful to visualize the JSI in the original
$\omega_r,\omega_s,\omega_i$ variables.   The structure of the JSI,
again akin to a narrow, tilted membrane unfortunately makes this a
difficult task. In Fig.~\ref{fig3Dstate} we have plotted the
function resulting from making each of the JSI frequency arguments
in turn equal to the degenerate frequency $\omega_p/3$, and
displayed each of the three resulting plots on the corresponding
plane in $\{\omega_s,\omega_r,\omega_i\}$ space. In
Fig.~\ref{fig3Dstate}, panel (a) shows a plot of the pump envelope
function $|\alpha(\omega_s+\omega_r+\omega_i)|^2$ [see
Eq.~(\ref{Eq:alphaGauss})], panel (b) shows a plot of the phase
matching function $|\phi(\omega_s,\omega_r,\omega_i)|^2$ [see
Eq.~(\ref{fphmat})], and panel (c) shows a plot of the JSI.  Note
that while the width of the phasematching function is proportional
to $1/L$, the width of the pump envelope function is proportional to
$\sigma$.   In order to make these plots graphically clear, we have
broadened each of the functions by selecting a fiber length of
$L/100$ and a pump bandwidth of $200 \sigma$, where $L$ and $\sigma$
are the values assumed for our frequency-degenerate source design.
While these are not meant to constitute physically feasible values,
they yield a three-dimensional appreciation of the ``membrane'',
except broadened,  in the generation frequencies space.  While in
Figs.~\ref{figest_rotado} and \ref{fig3Dstate} we have concentrated
on the frequency-degenerate source design, similar plots could be
made (but are not shown here) for the frequency non-degenerate
source design.

Besides the joint spectrum $|F(\omega_r,\omega_s,\omega_i)|^2$ of the emitted
photon triplets, we are also interested in the joint spectrum $I_2(\omega_r,\omega_s)$ of photon pairs
resulting from disregarding one of the photons in the triplet, and in the single-photon
spectrum $I_1(\omega_r)$ resulting from disregarding two of the photons in the triplet.  Functions
$I_2(\omega_r,\omega_s)$ and $I_1(\omega_r)$ are given by

\begin{eqnarray}\label{marg}
    I_2(\omega_r,\omega_s) &=&\int\!\!d\omega_i\left|F(\omega_r,\omega_s,\omega_i)\right|^2,
    \end{eqnarray}

\noindent and
\begin{eqnarray}
I_1(\omega_r)&=&\int\!\!d\omega_s\!\!\int\!\!d\omega_i \left|F(\omega_r,\omega_s,\omega_i)\right|^2.
\end{eqnarray}

Fig.~\ref{fig3Dstate}(d) shows a plot of the two-photon joint spectrum $I_2(\omega_r,\omega_s)$ which corresponds
to the three-photon joint spectrum of Fig.~\ref{fig3Dstate}(c).   Note that this two-photon joint spectrum
may be informally thought of as the shadow cast, on the $\{\omega_r,\omega_s\}$ plane, by the ``membrane'' discussed above.
Fig.~\ref{fig3Dstate}(e) shows a plot of the single-photon spectrum $I_1(\omega_r)$ which corresponds to the three-photon joint spectrum of Fig.~\ref{fig3Dstate}(c).

We now turn our attention to the case of frequency non-degenerate
TOSPDC, obtained by detuning the pump frequency while maintaining
other source parameters fixed, as discussed in the context of
Fig.~\ref{Fig:Gamma2}(b).   Let us assume that the pump frequency is
$\omega_p=2 \pi c/0.531\mu$m, i.e. with a $1$nm offset compared to
the value assumed for the frequency-degenerate source design, above.
As was studied in Fig.~\ref{Fig:Gamma2}(b), for a fixed idler
frequency, such a pump frequency offset leads to three distinct
phasematched frequencies for each of the the three TOSPDC modes:
$\omega_{r0}=2 \pi c/1.529\mu$m, $\omega_{s0}=2 \pi c/1.659\mu$m,
and $\omega_{i0}=2 \pi c/1.596\mu$m.  However, note that for
Fig.~\ref{Fig:Gamma2}(b), we have arbitrarily fixed the idler
frequency to the value $2 \pi c/1.596\mu$m.   In fact, we must
consider all idler frequencies, each leading to a plot similar to
Fig.~\ref{Fig:Gamma2}(b) with different $\omega_{r}$ and
$\omega_{s}$ values for a fixed $\omega_p$. Thus, in order for the
three emission modes to become spectrally distinct it is important
to spectrally filter the idler mode, so that in this specific
example only a small bandwidth centered at $\omega_{i0}$ is
retained.

\begin{figure}[ht]
\centering\includegraphics[width=8 cm]{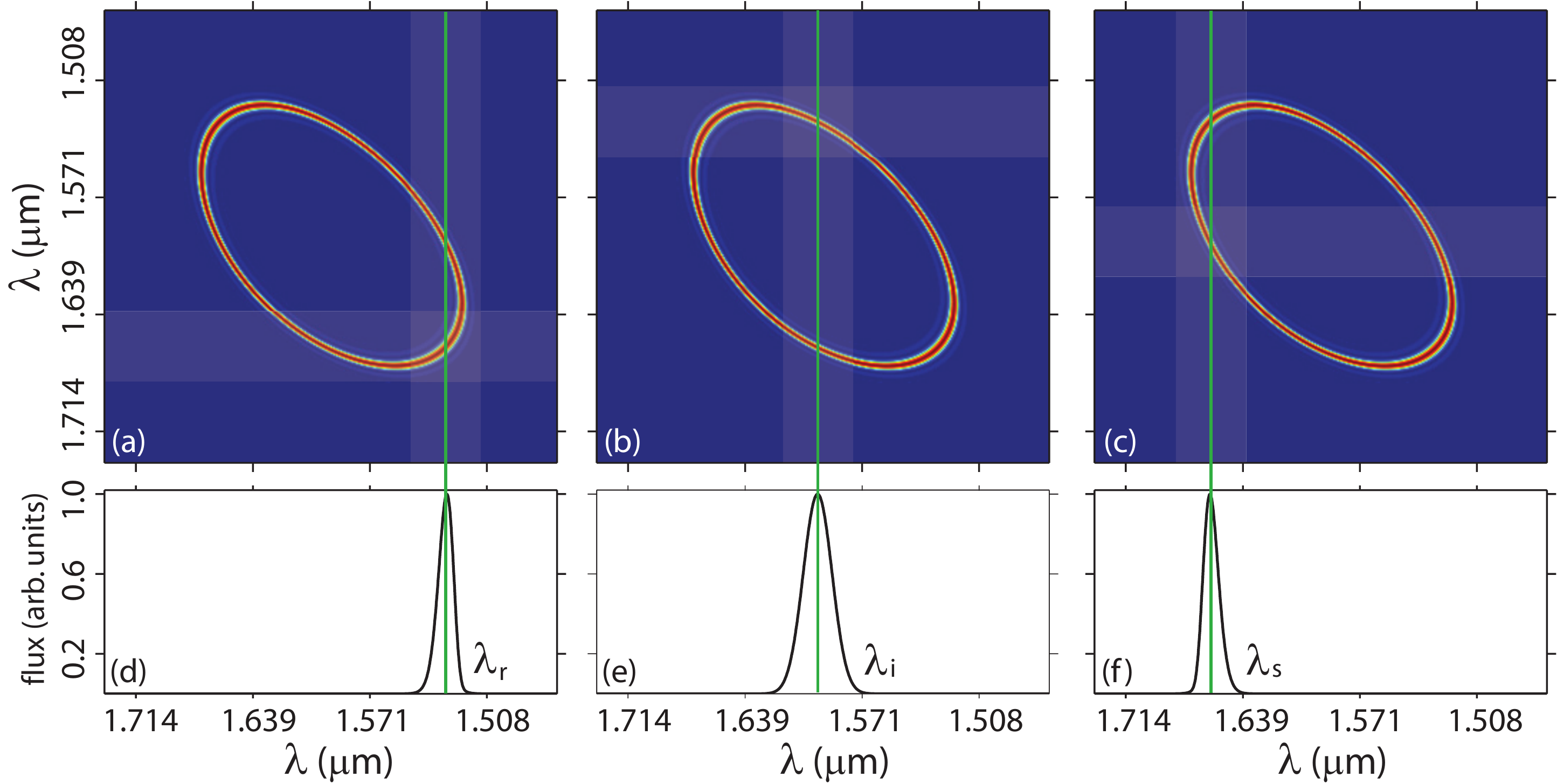}\caption{(color online)
Frequency non-degenerate TOSPDC photon-triplet state.
(a)-(c): Two-photon spectra obtained by integrating the JSI over each of the
three emission frequnecies in turn. The light shaded bands indicate
spectral filtering used.  (d)-(f) Single photon spectra for each of the
three emission modes, including the effect of spectral filtering.}
\label{Fig:3DstateND}
\end{figure}

The $1$nm offset in the pump wavelength from the previous paragraph
implies that the pump envelope function intersects the phasematching
function at a higher $\omega_{+}$ value (compared to that for the
degenerate source design) leading to a JSI which in the
$\{\nu_{A},\nu_{B},\nu_{+}\}$ space is a circular ring.  The
two-photon JSI's obtained by integrating the full JSI over each of
the TOSPDC frequencies in turn, $I_{2si}(\nu_s,\nu_i)$,
$I_{2ri}(\nu_r,\nu_i)$, and  $I_{2rs}(\nu_r,\nu_s)$ (where the
letter subscripts indicate the corresponding TOSPDC modes), then become
oblong rings, as shown in Fig.~\ref{Fig:3DstateND}(a)-(c).

By filtering each the three emission modes with Gaussian spectral
filters with bandwidth $\sigma_f=15$THz centered at each of the the
three selected phasematched frequencies, $\omega_{r0}$,
$\omega_{s0}$, and $\omega_{i0}$, we obtain the single-photon
spectra $I_{1r}(\nu)$, $I_{1s}(\nu)$, and $I_{1i}(\nu)$ (where the
letter subscript indicates the corresponding TOSPDC mode) shown in
Figs.~\ref{Fig:3DstateND}(d)-(f). Note that the spectral window
transmitted by each of these filters is indicated in
Fig.~\ref{Fig:3DstateND}(a)-(c) by a band with lighter shading.
Importantly, note that the three resulting generation modes do no
overlap each other.  This means that the photon triplets can then be
split into three separate modes deterministically by exploiting the
frequency differences among them.  This is achieved, however, at the
cost of a flux reduction resulting from the filters used.

In what follows, we
analyze the emitted flux for our two TOSPDC source designs, as a
function of several key experimental parameters.

\subsection{Emitted flux for specific TOSPDC source designs} \label{flux_design}

In this section we present numerical simulations of the expected
emitted flux, where possible comparing with results derived from
our analytic expressions in closed form.  In particular, we study
the dependence of the emitted flux vs certain key experimental
parameters: fiber length, pump power, and pump bandwidth.  We include
in this analysis our frequency-degenerate and frequency non-degenerate
designs of Sec.~\ref{designs}, as well as the pulsed-
and monochromatic-pump configurations.

We assume the following parameters: for the pulsed-pumped regime, a
bandwidth of $\sigma=23.5$GHz (which corresponds to a
Fourier-transform temporal duration of $100$ps), except in
Sec.~\ref{bandw} where we analyze the emitted flux vs $\sigma$
dependence; a fiber length of $L=10$cm except in Sec. \ref{Long},
where we discuss the  emitted flux vs fiber length dependence; an
average pump power $p=200$mW except in Sec. \ref{pot} where analyze
emitted flux vs pump power dependence.


\subsubsection{Pump bandwidth dependence} \label{bandw}

In this subsection we study the dependence of the emitted flux for our two source designs
on the pump bandwidth, while
maintaining the energy per pump pulse constant.
Note that as $\sigma$ varies, the temporal duration varies, and
consequently the peak power varies too. We evaluate the emitted flux
for a pump bandwidth $\sigma$ range $11.77-117.7$GHz (or a
Fourier-transform-limited temporal duration range $20-200$ps).

\begin{figure}[ht]
\centering\includegraphics[width=6 cm]{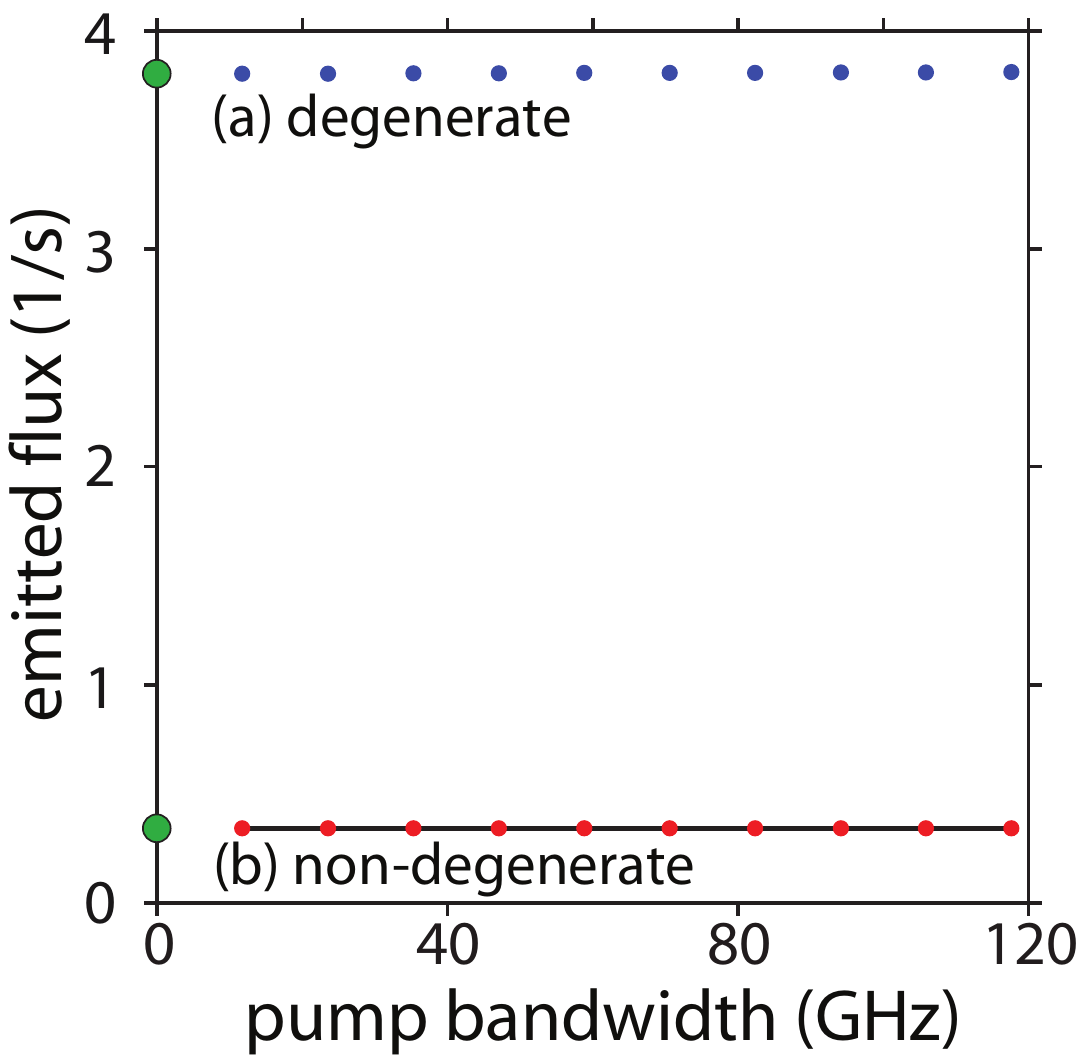}{0.5}\caption{(color online) Emitted flux
as a function of the pump bandwidth, for the following cases: (a)
Frequency-degenerate TOSPDC source, evaluated from
Eq.~(\ref{Eq:NumFot}) (blue dots); (b) Frequency non-degenerate
TOSPDC source, evaluated from Eq.~(\ref{Eq:NumFot}) (red dots), and
frequency non-degenerate TOSPDC from the closed analytic expression,
i.e. Eq.~(\ref{Napprox}) (black solid line). Values obtained for the
monochromatic pump limit, through Eq.~(\ref{Ncw}), are indicated by
green dots.
 } \label{fluxSigma}
\end{figure}

For both source designs, the emitted flux is obtained by numerical
evaluation of Eq.~(\ref{Eq:NumFot}). Results are shown in
Fig.~\ref{fluxSigma} by blue dots for the degenerate case, and by
red dots for the non-degenerate case. We have also obtained from
Eq.~(\ref{Ncw}) the emitted flux in the monochromatic-pump limit,
shown in Fig.~\ref{fluxSigma} by green dots.   It is graphically
clear that the emitted flux values for $\sigma \neq 0$ [calculated
from Eq.~(\ref{Eq:NumFot})] approach the corresponding values in the
monochromatic-pump limit [calculated from Eq.~(\ref{Ncw})].
Additionally, for our TOSPDC non-degenerate source, we evaluate the
emitted flux from the analytical expression given in
Eq.~(\ref{Napprox}), and the corresponding results are shown in
Fig.~\ref{fluxSigma} by the black-solid line. As can be seen, the
agreement between numerical and analytical results is excellent,
indicating that the linear approximation on which the analytic
results are based is in fact a good approximation.  As was discussed
in Sec.~\ref{secAnali}, this approximation fails for the
frequency-degenerate case.

As is clear from Fig.~\ref{fluxSigma}, the TOSPDC emitted flux (and
therefore the conversion efficiency) remains constant vs pump bandwidth over the full
range of pump bandwidths considered, for both the degenerate and
non-degenerate photon-triplet sources. For this reason, in the case
of TOSPDC, no difference is expected in the emitted flux, between
the monochromatic- and pulsed-pump regimes (while maintaining the
average pump power constant).

Note also that the frequency-degenerate
source is significantly brighter than the frequency non-degenerate source;
the reason for this is that at $\omega_r=\omega_s=\omega_i=\omega_p/3$, the
perfect phasematching contour and the energy conservation contour are
tangent to each other, leading to a greater emission bandwidth.
Our results yield a source brightness of $N=3.80\, \mbox{triplets/s}$ for the degenerate
source, and a value of $N=0.34\, \mbox{triplets/s}$ for the
non-degenerate TOSPDC source. It should be noted, however,
that for the frequency-degenerate case, photon triplets may be split only
non-deterministically so that the actual usuable source brightness may be lower
that our results would indicate.

\subsubsection{Fiber length dependence}\label{Long}

We now turn our attention to the fiber-length dependence of the
emitted flux from both the degenerate and non-degenerate TOSPDC
sources, while maintaining other source parameters fixed. For this
analysis we vary the fiber length from $1$ to $10$cm. Note that a
recent experimental work shows that it is possible to obtain a
uniform-radius fiber taper of $\sim 445$nm radius over a length of
$9$cm \cite{lsaval04}.

\begin{figure}[ht]
\centering\includegraphics[width=6 cm]{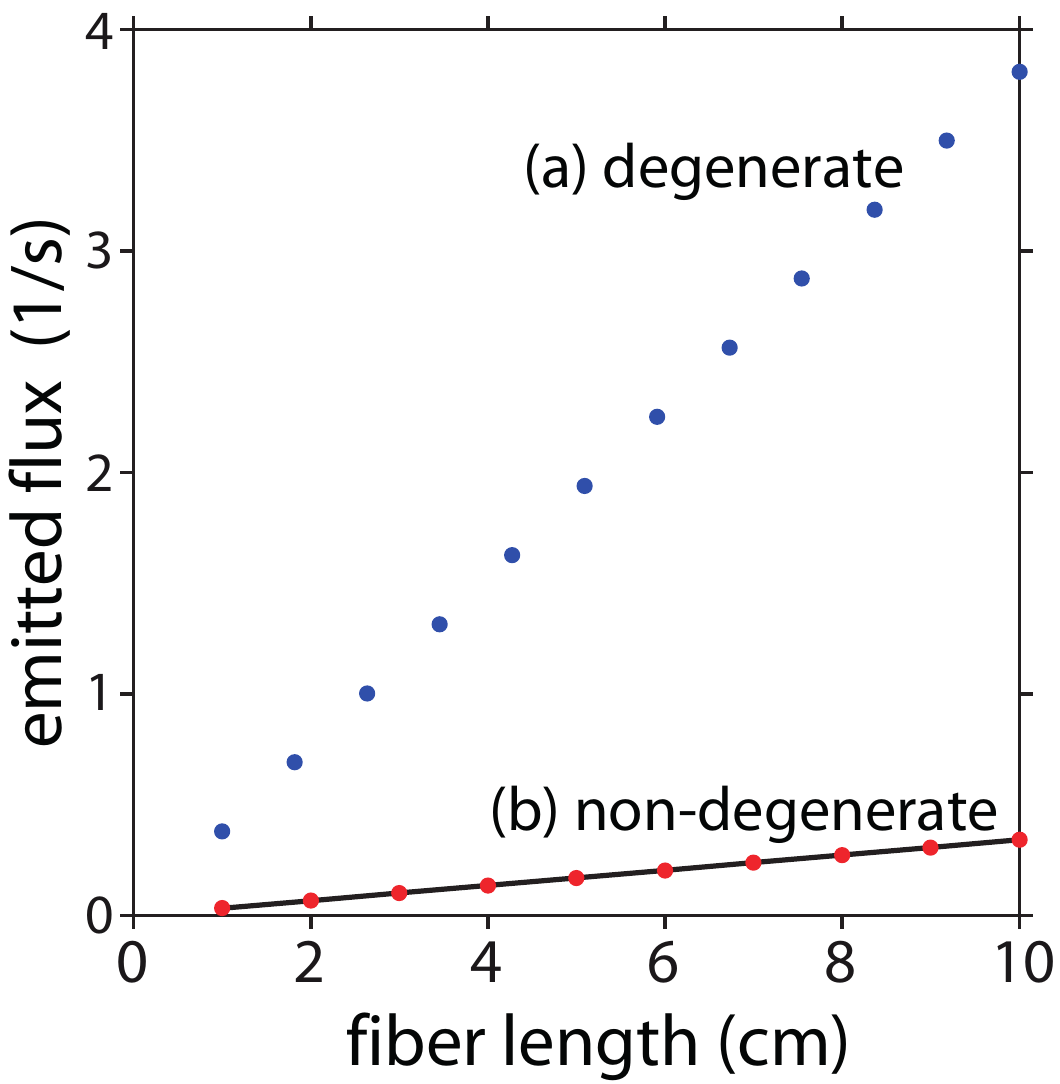}{0.5}\caption{(color online) Emitted flux as
a function of the fiber length, for the following cases:
(a) Frequency-degenerate TOSPDC source, evaluated
from Eq.~(\ref{Eq:NumFot}) (blue dots); (b) Frequency non-degenerate
TOSPDC source, evaluated
from Eq.~(\ref{Eq:NumFot}) (red dots), and frequency non-degenerate TOSPDC from closed analytic expression, i.e. Eq.~(\ref{Napprox}) (black solid line).}
\label{fluxL}
\end{figure}

The results obtained by numerical evaluation of
Eq.~(\ref{Eq:NumFot}) are shown in Fig.~\ref{fluxL} by
blue dots (degenerate case), and by red dots
(non-degenerate case). We have also evaluated from  Eq.~(\ref{Ncw})
the emitted flux obtained in the monochromatic-pump limit. However,
because the emitted flux is constant with respect to the pump
bandwidth (for the experimental parameters assumed here), the
values obtained overlap those resulting from Eq.~(\ref{Eq:NumFot}), for the pulsed pump regime.
Additionally, for the TOSPDC non-degenerate source, we evaluate the
emitted flux from the analytical expression given in
Eq.~(\ref{Napprox}). The corresponding results, which are shown
graphically in Fig.~\ref{fluxL} by the black-solid line, are in
excellent agreement with those obtained from Eq.~(\ref{Eq:NumFot}).

Note that for the fiber length range considered, the emitted flux
exhibits a linear dependence on $L$ for both the frequency-degenerate
and the non-degenerate photon-triplet sources. However, it should be
noted that there are conditions for which $N$ has a nonlinear
dependence on the fiber length. For example, as it was discussed in
the Sec.~\ref{secAnali}, for $L\ll L_0$ the emitted flux varies quadratically with the
fiber length. For the longest fiber considered here ($L=10$cm), the
TOSPDC emitted flux for the degenerate source is $N=3.80\,
\mbox{triplets/s}$ and $N=0.34\, \mbox{triplets/s}$ for the
non-degenerate source.

\subsubsection{Pump power dependence}\label{pot}

We now turn our attention to the pump-power dependence of the
emitted flux for the two TOSPDC sources, while maintaining the pump
bandwidth and other source parameters fixed. We compute the emitted
flux as a function of the average pump power, which is varied
between $1$ and $200$ mW.

\begin{figure}[ht]
\centering\includegraphics[width=6 cm]{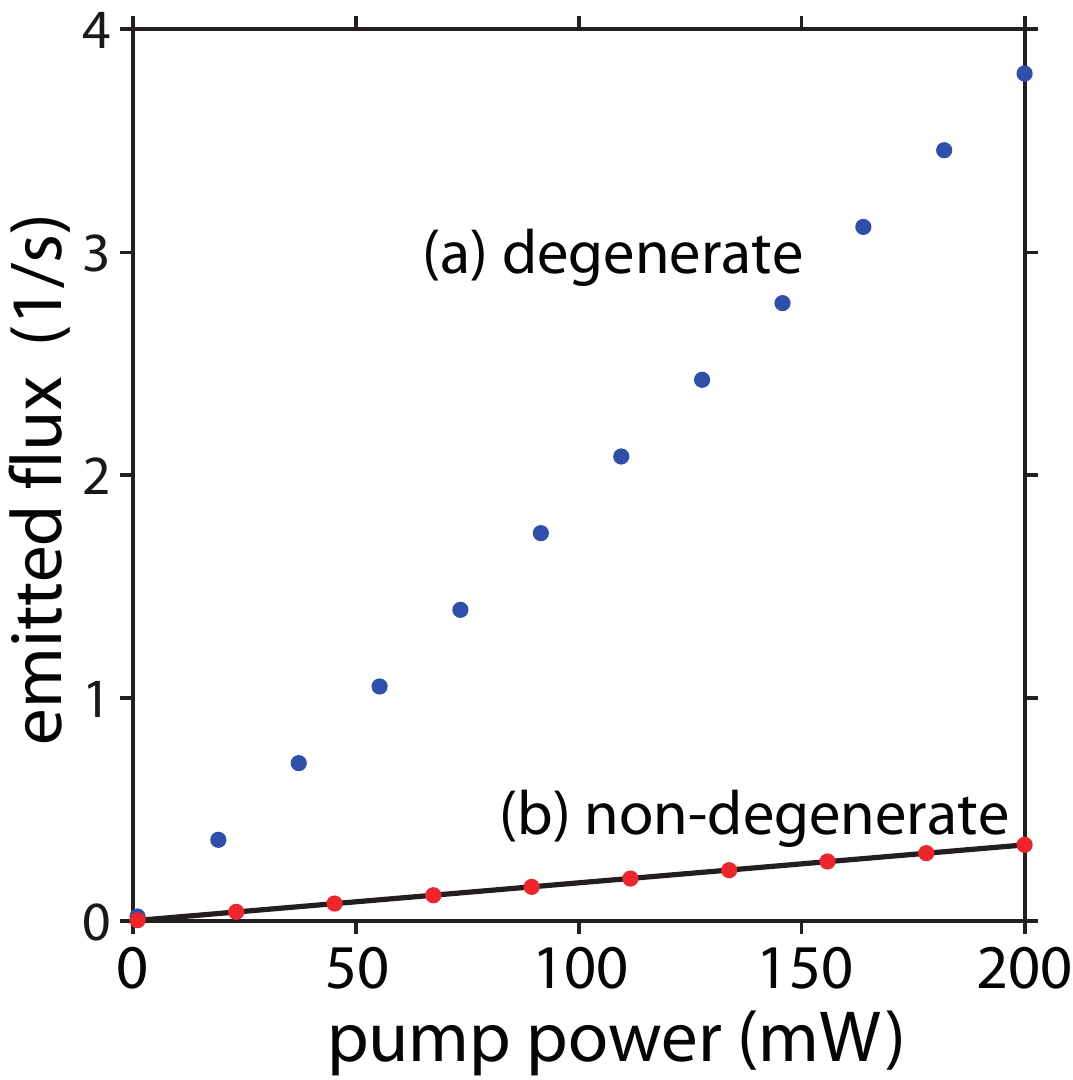}{0.5}\caption{ (color online) Emitted flux as
a function of the pump power, for the following cases:
(a) Frequency-degenerate TOSPDC source, evaluated
from Eq.~(\ref{Eq:NumFot}) (blue dots); (b) Frequency non-degenerate
TOSPDC source, evaluated
from Eq.~(\ref{Eq:NumFot}) (red dots), and from closed analytic expression, i.e. Eq.~(\ref{Napprox}) (black solid line).} \label{fluxPav}
\end{figure}

In Fig.~\ref{fluxPav} we present, for the two proposed source
designs, plots of $N$ vs $p$, which were obtained numerically from
the expression in Eq.~(\ref{Eq:NumFot}). The blue dots
correspond to the degenerate case, while the red dots correspond to
the non-degenerate case. Plots of the emitted flux
obtained in the monochromatic-pump limit are not shown in
Fig.~\ref{fluxPav}, because they overlap results obtained from
Eq.~(\ref{Eq:NumFot}) for the pulsed-pump regime. Additionally, for
our TOSPDC non-degenerate source, we evaluate the emitted flux from
the analytical expression given in Eq.~(\ref{Napprox}). Corresponding
results are shown in Fig.~\ref{fluxPav} by the black-solid line. As
can be seen, the agreement between the numerical and
the analytical results is excellent.

As can be seen in  Fig.~\ref{fluxPav}, for both sources the emitted
flux depends linearly on the pump average power, which implies that
the TOSPDC conversion efficiency is constant with respect to this
parameter [see Eqs.~(\ref{Eq:conversioneff1}) and (\ref{eficw})].
This behavior should be contrasted with the SFWM process, for which
the conversion efficiency is linear with respect to the pump power
\cite{garay10}. Note that the process of TOSPDC has important
similarities with the process of SPDC; in both cases, the conversion
efficiency is constant with respect to the pump power and to the
pump bandwidth (within the phasematching bandwidth).

At the highest average pump power considered here ($p=200$ mW), the
TOSPDC emitted flux for the degenerate source is $N=3.80
\mbox{triplets/s}$.

\section{Conclusions}

In this paper we have studied the third-order spontaneous parametric
downconversion process, including both the frequency-degenerate and
frequency non-degenerate cases, implemented in thin optical fibers.
We have based our analysis on a configuration introduced in an
earlier paper from our group (see Ref.~\cite{corona11}), in which
the pump and the generated modes propagate in different fiber modes,
with the objective of attaining phasematching.   In this paper we
study the emitted photon-triplet TOSPDC states, and present two
different ways to visualize this state.  We present an analysis of
the photon-triplet emission flux, which leads to expressions in
integral form which for frequency non-degenerate TOSPDC are taken to
closed analytic form under certain approximations.  We show plots of
the emitted flux as a function of several key parameters, obtained
through numerical evaluation of our full expressions, where possible
comparing with results derived from our closed analytic expressions.
We also analyze the TOSPDC phasematching characteristics of thin
optical fibers, in particular as a function of the fiber radius and
the pump frequency.   We hope that this paper will be useful as the
basis for the practical implementation of photon triplet sources
based on third-order spontaneous parametric downconversion.

\begin{acknowledgements}
This work was supported in part by CONACYT, Mexico,  by DGAPA, UNAM
and by FONCICYT project 94142.
\end{acknowledgements}


\begin{thebibliography}{99}
\bibitem {burnham70} D. C. Burnham and D. L. Weinberg, Phys. Rev. Lett. \textbf{25}, 84, 135-179 (1970).

\bibitem {fiorentino02}M. Fiorentino, P. L. Voss, J. E. Sharping, and P. Kumar, IEEE Photonics Technol. Lett. \textbf{14}, 983 (2002).

\bibitem {chekhova05}M. V. Chekhova, O. A. Ivanova, V. Berardi, and A. Garuccio, Phys. Rev. A \textbf{72}, 023818 (2005).

\bibitem{hnilo05}A. A. Hnilo, Phys. Rev. A \textbf{71}, 033820
(2005).

\bibitem {felbinger98}T. Felbinger, S. Schiller, and J. Mlynek, Phys. Rev. Lett. \textbf{80}, 492--495 (1998).

\bibitem{bencheikh07} K. Bencheikh, F. Gravier, J. Douady, A.
Levenson, and B. Boulanger, C. R. Physique \textbf{8}, 206--220
(2007).

\bibitem{banaszek97}K. Banaszek, and P.  L. Knight, Phys. Rev. A \textbf{55}, 2368--2375 (1997).

\bibitem {douady04}J. Douady, and B. Boulanger, Opt. Lett. \textbf{29}, 2798 (2004).

\bibitem {Sliwa03}C. \'Sliwa, and K. Banaszek, Phys. Rev. A \textbf{67}, 030101(R) (2003).

\bibitem {Wagenknecht10} C. Wagenknecht, C.-M. Li, A. Reingruber, X.-H. Bao, A. Goebel, Y.-A. Chen, Q. Zhang, K. Chen, and J.-W. Pan, Nature Photon. \textbf{4}, 549--552 (2010).

\bibitem {Barz10}S. Barz, G. Cronenberg, A. Zeilinger, and P.
Walther, Nature Photon. \textbf{4}, 553--556 (2010).

\bibitem {greenberger90} D. M. Greenberger, M. A. Horne, A. Shimony, and A. Zeilinger, A. J. Phys. \textbf{58}, 1131 (1990).

\bibitem{bouwmeester03} D. Bouwmeester, J.-W. Pan, M. Daniell, H. Weinfurter, and A.
Zeilinger, Phys. Rev. Lett. \textbf{82}, 1345-1349 (1999).

\bibitem {persson04} J. Persson, T. Aichele, V. Zwiller, L. Samuelson, and O. Benson, Phys. Rev. B \textbf{69}, 233314 (2004).

\bibitem {keller98}T. E. Keller, M. H. Rubin, and Y. Shih, Phys. Rev. A \textbf{57}, 2076 (1998).

\bibitem {hubel10} H. H\"ubel, D. R. Hamel, A. Fedrizzi, S. Ramelow, K.J. Resch, and T. Jennewein, Nature \textbf{466}, 601 (2010).

\bibitem {antonosyan11}D. A. Antonosyan, T. V. Gevorgyan, and G. Yu. Kryuchkyan, Phys.
Rev. A \textbf{83}, 043807 (2011).

\bibitem {rarity99} J. G. Rarity and P. R. Tapster, Phys. Rev. A 59, R35 (1999).

\bibitem{corona11} M. Corona, K. Garay-Palmett, and A. B.
U'Ren, Opt. Lett. \textbf{36}, 190--192 (2011).

\bibitem {mandel}L. Mandel and E. Wolf, \textit{Optical Coherence and Quantum Optics}\/ (Cambridge University Press, 1995).

\bibitem {agrawal07}G. P. Agrawal, \textit{Nonlinear Fiber Optics, 4th Ed.} (Elsevier, 2007).

\bibitem{garay10} K. Garay-Palmett, A. B. U'Ren, and R. Rangel-Rojo. Phys. Rev. A 82, 043809 (2010).

\bibitem{kolevatova03} O. A. Kolevatova, A. N. Naumonv, and A. M.
Zheltikov, Laser Phys. \textbf{13}, 1040--1045 (2003).

\bibitem{grubsky07} V. Grubsky and J. Feinberg, Optics Comm. 274, pp. 447-450 (2007).

\bibitem{tong03} L. Tong, R. G. Gattas, J. A. Ashcom, S. He, J. Lou, M. Shen, I. Maxwell
and E. Mazur, Nature \textbf{426}, 816--819 (2003).

\bibitem{lsaval04} S. Leon-Saval, T. Birks, W. Wadsworth, P. St. J. Russell, and M. Mason, Opt. Express \textbf{12}, 2864 (2004).

\bibitem{brambilla10} G. Brambilla, J. Opt. \textbf{12}, 043001 (2010).

%




\end{thebibliography}
\end{document}